\documentclass[twocolumn]{aastex631}
\usepackage{hyperref}

\usepackage{float}
\usepackage{verbatim}
\usepackage{natbib}
\usepackage{amsmath}

\usepackage{xcolor}

\newcommand{\ron}[1]{{\textcolor{red}{#1}}}

\begin{document}

\title{TIC 114936199: A Quadruple Star System with a 12-day Outer Orbit Eclipse}

\correspondingauthor{Brian P. Powell}
\email{brian.p.powell@nasa.gov}

\author[0000-0003-0501-2636]{Brian P. Powell}
\affiliation{NASA Goddard Space Flight Center, 8800 Greenbelt Road, Greenbelt, MD 20771, USA}
%
\author[0000-0003-3182-5569]{Saul A. Rappaport}
\affiliation{Department of Physics, Kavli Institute for Astrophysics and Space Research, M.I.T., Cambridge, MA 02139, USA}
%
\author[0000-0002-8806-496X]{Tam\'as Borkovits}
\affiliation{Baja Astronomical Observatory of University of Szeged, H-6500 Baja, Szegedi út, Kt. 766, Hungary}
\affiliation{ELKH-SZTE Stellar Astrophysics Research Group, H-6500 Baja, Szegedi \'ut, Kt. 766, Hungary}
\affiliation{Konkoly Observatory, Research Centre for Astronomy and Earth Sciences, H-1121 Budapest, Konkoly Thege Miklós út 15-17, Hungary}
\affiliation{ELTE Gothard Astrophysical Observatory, H-9700 Szombathely, Szent Imre h. u. 112, Hungary}
\affiliation{MTA-ELTE Exoplanet Research Group, H-9700 Szombathely, Szent Imre h. u. 112, Hungary}
%
%
\author[0000-0001-9786-1031]{Veselin~B.~Kostov}
\affiliation{NASA Goddard Space Flight Center, 8800 Greenbelt Road, Greenbelt, MD 20771, USA}
\affiliation{SETI Institute, 189 Bernardo Ave, Suite 200, Mountain View, CA 94043, USA}
\affiliation{GSFC Sellers Exoplanet Environments Collaboration}
%
\author[0000-0002-5286-0251]{Guillermo Torres}
\affiliation{Center for Astrophysics $\vert$ Harvard \& Smithsonian, 60 Garden Street, Cambridge, MA 02138, USA}
%
\author[0000-0002-7778-3117]{Rahul Jayaraman}
\affiliation{Department of Physics, Kavli Institute for Astrophysics and Space Research, M.I.T., Cambridge, MA 02139, USA}
%
\author[0000-0001-9911-7388]{David W. Latham}
\affiliation{Center for Astrophysics $\vert$ Harvard \& Smithsonian, 60 Garden Street, Cambridge, MA 02138, USA}
\author[0000-0002-1330-1318]{Hana Ku\v{c}\'akov\'a}
\affiliation{Astronomical Institute of the Czech Academy of Sciences, Fri\v{c}ova 298, CZ-251 65 Ond\v{r}ejov, Czech Republic}
\affiliation{Research Centre for Theoretical Physics and Astrophysics, Institute of Physics, Silesian University, Bezru\v{c}ovo n\'am. 13, CZ-746 01 Opava, Czech Republic}
\affiliation{Astronomical Institute, Faculty of Mathematics and Physics, Charles University Prague, CZ-180 00 Praha 8, V Hole\v{s}ovi\@{c}k\'ach 2, Czech Republic}
%
\author[0000-0001-9483-2016]{Zolt\'an Garai}
\affiliation{ELTE Gothard Astrophysical Observatory, H-9700 Szombathely, Szent Imre h. u. 112, Hungary}
\affiliation{MTA-ELTE Exoplanet Research Group, H-9700 Szombathely, Szent Imre h. u. 112, Hungary}
\affiliation{Astronomical Institute, Slovak Academy of Sciences, 05960 Tatransk\'a 
Lomnica, Slovakia}
%
\author[0000-0003-3599-516X]{Theodor Pribulla}
\affiliation{Astronomical Institute, Slovak Academy of Sciences, 05960 Tatransk\'a 
Lomnica, Slovakia}
%
\author[0000-0001-7246-5438]{Andrew Vanderburg}
\affiliation{Department of Physics, Kavli Institute for Astrophysics and Space Research, M.I.T., Cambridge, MA 02139, USA}
%
\author[0000-0002-0493-1342]{Ethan Kruse}
\affiliation{NASA Goddard Space Flight Center, 8800 Greenbelt Road, Greenbelt, MD 20771, USA}
\affiliation{Department of Astronomy, University of Maryland, College Park, MD 20742, USA}
%
\author[0000-0001-7139-2724]{Thomas~Barclay}
\affiliation{NASA Goddard Space Flight Center, 8800 Greenbelt Road, Greenbelt, MD 20771, USA}
\affiliation{University of Maryland, Baltimore County, 1000 Hilltop Circle,
Baltimore, MD 21250, USA}

\author[0000-0001-8472-2219]{Greg Olmschenk}
\affiliation{NASA Goddard Space Flight Center, 8800 Greenbelt Road, Greenbelt, MD 20771, USA}
\affiliation{Universities Space Research Association, 7178 Columbia Gateway Drive, Columbia, MD 21046}
%
\author[0000-0002-2607-138X]{Martti~H.~K.~Kristiansen}
\affil{Brorfelde Observatory, Observator Gyldenkernes Vej 7, DK-4340 T\o{}ll\o{}se, Denmark}

\author[0000-0002-5665-1879]{Robert Gagliano}
\affiliation{Amateur Astronomer, Glendale, AZ 85308}
%
\author[0000-0003-3988-3245]{Thomas L. Jacobs}
\affiliation{Amateur Astronomer, 12812 SE 69th Place, Bellevue, WA 98006}
%

\author[0000-0002-8527-2114]{Daryll M. LaCourse}
\affiliation{Amateur Astronomer, 7507 52nd Place NE Marysville, WA 98270}
%
\author{Mark Omohundro}
\affiliation{Citizen Scientist, c/o Zooniverse, Department of Physics, University of Oxford, Denys Wilkinson Building, Keble Road, Oxford, OX13RH, UK}
%
\author[0000-0002-1637-2189]{Hans M. Schwengeler}
\affiliation{Citizen Scientist, Planet Hunter, Bottmingen, Switzerland}
%
\author[0000-0002-0654-4442]{Ivan A. Terentev}
\affiliation{Citizen Scientist, Planet Hunter, Petrozavodsk, Russia}
%
\author[0000-0002-5034-0949]{Allan R. Schmitt}
\affiliation{Citizen Scientist, 616 W. 53rd. St., Apt. 101, Minneapolis, MN 55419, USA}

%
\received{3 August 2022}
\revised{9 August 2022}
\accepted{10 August 2022}

\begin{abstract}
We report the discovery with {\it TESS} of a remarkable quadruple star system with a 2+1+1 configuration.  The two unique characteristics of this system are that (i) the inner eclipsing binary (stars Aa and Ab) eclipses the star in the outermost orbit (star C), and (ii) these outer 4th body eclipses last for $\sim$12 days, the longest of any such system known.  The three orbital periods are $\sim$3.3 days, $\sim$51 days, and $\sim$2100 days.  The extremely long duration of the outer eclipses is due to the fact that star B slows binary A down on the sky relative to star C.  We combine {\it TESS} photometric data, ground-based photometric observations, eclipse timing points, radial velocity measurements, the composite spectral energy distribution, and stellar isochones in a spectro-photodynamical analysis to deduce all of the basic properties of the four stars (mass, radius, $T_{\rm eff}$, and age), as well as the orbital parameters for all three orbits.  The four masses are $M_{\rm Aa} =0.382$\,M$_\odot$, $M_{\rm Ab} =0.300$\,M$_\odot$, $M_{\rm B} =0.540$\,M$_\odot$ and $M_{\rm C} =0.615$\,M$_\odot$, with a typical uncertainty of 0.015 M$_\odot$. 

\end{abstract}

\keywords{Multiple Stars -- Eclipsing Binary Stars}

\section{Introduction}\label{sec:intro}

The Transiting Exoplanet Survey Satellite ({\em TESS}) mission \citep{Ricker14} has allowed for the identification of many spectacular multiple star systems through eclipses in the Full Frame Image (FFI) light curves.  For example, the near-perfectly coplanar and circular triple star system TIC 278825952, discovered by \citet{2020MNRAS.498.6034M}, which contains an eclipsing binary (EB) that causes both eclipses and occulations of the tertiary star.  Discoveries of many other interesting triply eclipsing triple star systems with {\em TESS} have been made by \citet{2020MNRAS.496.4624B,2022MNRAS.510.1352B} and \citet{2022MNRAS.513.4341R}.  The quadruple star system TIC 454140642 \citep{2021ApJ...917...93K} is also of note as a compact and nearly coplanar system of two EBs consisting of similar mass stars, coming within 1 AU of each other during the course of their 432 day outer orbit. BG Ind, a well-known, nearby EB, was discovered by \citet{2021MNRAS.503.3759B} to be a quadruple system due to the presence of a second set of eclipses in the {\em TESS} light curve. Additionally, \citet{2022ApJS..259...66K} and \citet{2022arXiv220503934Z} have produced catalogs of quadruple star systems consisting of pairs of likely gravitationally bound EBs discovered with {\em TESS}, together more than doubling the number of such systems that are known.  Furthermore, the sextuple star system TIC 168789840, discovered by \citet{2021AJ....161..162P}, consists of three EBs in a (2+2)+2 hierarchical configuration. For further discussion of eclipsing multiple star systems discovered with {\em TESS} we refer the reader to the reviews of \citet{2021Univ....7..369S} and \citet{2022Galax..10....9B}.

Wide triple star systems with outer separations of $\gtrsim 10$ AU are the most common among triples (see, e.g., \citealt{2021Univ....7..352T} or \citealt{2018ApJS..235....6T}).  Compact triple systems, defined here as having an outer orbital period of less than a few years, are becoming more common with the advent of spaced based photometric sky surveys such as CoRoT \citep{2009A&A...506..411A}, {\it Kepler} \citep{2010Sci...327..977B}, and {\it TESS}.  Perhaps the most fascinating of the compact triples are those that exhibit third-body eclipses (see, e.g., \citealt{2016MNRAS.455.4136B,2022MNRAS.510.1352B,2022MNRAS.513.4341R}).  But, as eclipses and multi-stellar eclipsing systems become more well known and studied, we have noticed a glaring absence of quadruple star systems with outer orbit eclipses, including among the many quadruples recently discovered by \citet{2022ApJS..259...66K} and \citet{2022arXiv220503934Z}.  However, two systems identified in the {\em Kepler} mission, KIC 5255552 and KIC 285696, have been suggested to demonstrate such eclipses.

KIC 5255552 was first identified as an EB by \citet{2011AJ....141...83P}. \citet{2015MNRAS.448..946B} discovered additional eclipses, analyzing the system as part of a study on eclipsing triple star systems which exhibit eclipse timing variations (ETVs).  It was further analyzed by \citet{2016MNRAS.455.4136B}, who solved for the system parameters including a third-body orbit at a period of ~862 days that was responsible for the three observed sets of extraneous eclipses. \citet{2018A&A...610A..72Z} identified a fourth set of eclipses and also suggested the presence of a fourth body in the system, though at a separate period. \citet{2020MNRAS.498.4356G} first suggested that the two additional stars in the system may be in a non-eclipsing binary, which is then involved in a transit event with the EB.  They also present the possibility of a fifth star in the system, although no physical model of this system as a quadruple was provided by either \citet{2018A&A...610A..72Z} or \citet{2020MNRAS.498.4356G}.

The so-called ``impossible triple'' \citep{2014MNRAS.445..309M}, KIC 2856960, contains an eclipsing binary with a period of $\sim$0.258 days with an additional complex set of third-body eclipsing events containing multiple eclipses. These occur over a duration of $\sim$1.3 days and repeat at a period of $\sim$204 days.  Like KIC 5255552, it was initially identified as an EB by \citet{2011AJ....141...83P}, with \citet{2012A&A...545L...4A} discovering additional, irregular eclipses and initially theorizing that it was a triple star system.  \citet{2013ApJ...763...74L} conducted an extensive analysis of the system using the ETV curve of the EB, and concluded that it was indeed a triple. \citet{2014MNRAS.445..309M} made the first attempt to model the third body events and deemed the system ``impossible'' as they found a triple star system could not satisfy Kepler's laws under the constraints of a configuration required to create the third body eclipses.  As such, the authors conclude that KIC 2856960 must be a quadruple star system, though a physical model of the system capable of producing the third body eclipses has yet to be published.

There is certainly an explainable reason for the lack of known quadruple systems with outer orbit eclipses, which is that, empirically, the outer orbit of quadruple star systems tend to be rather long in comparison to that of a triple consisting of stars of similar mass.  There are a substantial number of triples with periods in the range of 33 to 200 days (see e.g. Figure 1 of \citealt{2022MNRAS.510.1352B}), while in the case of quadruples, the shortest outer period known is $\sim$355 days \citep{2020MNRAS.494..178P}. As the period of the outer orbit increases, the probability of an eclipse on that orbit decreases roughly as $P^{-2/3}$. This probability is further reduced by the observational interval divided by the period.

TIC 114936199, with basic stellar parameters shown in Table \ref{tab:parameters}, is an exceedingly rare quadruple star system that exhibits eclipses on the outer orbit.  Not only does TIC 114936199 demonstrate an outer orbit eclipse, but the full duration of the transit event is an impressive $\sim$12 days.  This is so long, in fact, that it requires the relative motions of the inner and outer orbits to have a near-zero relative velocity for several days.

The structure of this paper is as follows:  In Section \ref{sec:photometric}, we present the distinctive TIC 114936199 light curve. Then, we discuss clues to the system configuration provided by the light curve in Section \ref{sec:clues}. Our initial estimates of the orbits of the system are shown in Section \ref{sec:orbits}, followed by spectroscopic observations in Section \ref{sec:spec}. We discuss our modeling process for the system in Section \ref{sec:system_model} and our final model in Section \ref{sec:mcmc}.  The fit to the Spectral Energy Distribution (SED) is shown in Section \ref{sec:sed}, followed by a discussion in Section \ref{sec:discussion}. We note the possibility of a fifth star in the system in Section \ref{sec:quintuple}, and summarize our findings in Section \ref{sec:summary}.

\begin{deluxetable}{l r r r }
\tabletypesize{\small}
\tablecaption{Stellar parameters of TIC 114936199.\label{tab:parameters}}
\tablewidth{0pt}
\tablehead{
\colhead{Parameter} & \colhead{Value} & \colhead{Error} &\colhead{Source}
}
\startdata
\multicolumn{4}{l}{\bf Identifying Information} \\
\hline
TIC ID & 114936199 & & TIC \\
{\it Gaia} ID & 4533878463922421376 & & {\it Gaia} DR3 \\
2MASS ID & 	19012862+2455338 & & TIC \\
ALLWISE ID & J190128.62+245534.1 & & TIC \\
RA  (hh:mm:ss) & 19:01:28.627 &   & TIC \\
Dec (dd:mm:ss) & +24:55:33.86 &   & TIC\\
Parallax (mas) & 7.694 & 0.081 & {\it Gaia} DR3 \\
PMRA (mas/yr) & 5.824 & 0.074 & {\it Gaia} DR3 \\
PMDEC (mas/yr) & 32.188 & 0.109 & {\it Gaia} DR3 \\
RUWE & 6.124 & & {\it Gaia} DR3 \\
\hline
\\
\multicolumn{4}{l}{\bf Photometric Magnitudes$^a$} \\
\hline
$T$  & 11.992 & 0.006 & TIC \\
$B$  & 15.793 & 0.162 & TIC \\
$V$  & 13.517 & 0.069 & TIC \\
$G$  & 	12.911 & 0.001  & TIC \\
$J$  & 10.557 & 0.023 & TIC \\
$H$  & 9.932 & 0.024 & TIC \\
$K$  & 9.733 & 0.02 & TIC \\
$W1$  & 9.62 & 0.023 & TIC \\
$W2$  & 9.568 & 0.02 & TIC \\
$W3$  & 9.402 & 0.033 & TIC \\
$W4$  & 9.186 & 0.538 & TIC \\
\enddata
{Notes. (a) $T$ = {\em TESS}, $B$ = Johnson B, $V$ = Johnson V, $G$ = {\em Gaia}, $J$ = 2MASS J, $H$ = 2MASS H, $K$ = 2MASS K, $W1$ = WISE W1, $W2$ = WISE W2, $W3$ = WISE W3, $W4$ = WISE W4.}
\end{deluxetable} 

\section{Photometric Observations}
\label{sec:photometric}

\begin{figure}[h]
    \centering
    \includegraphics[width=1.0\linewidth]{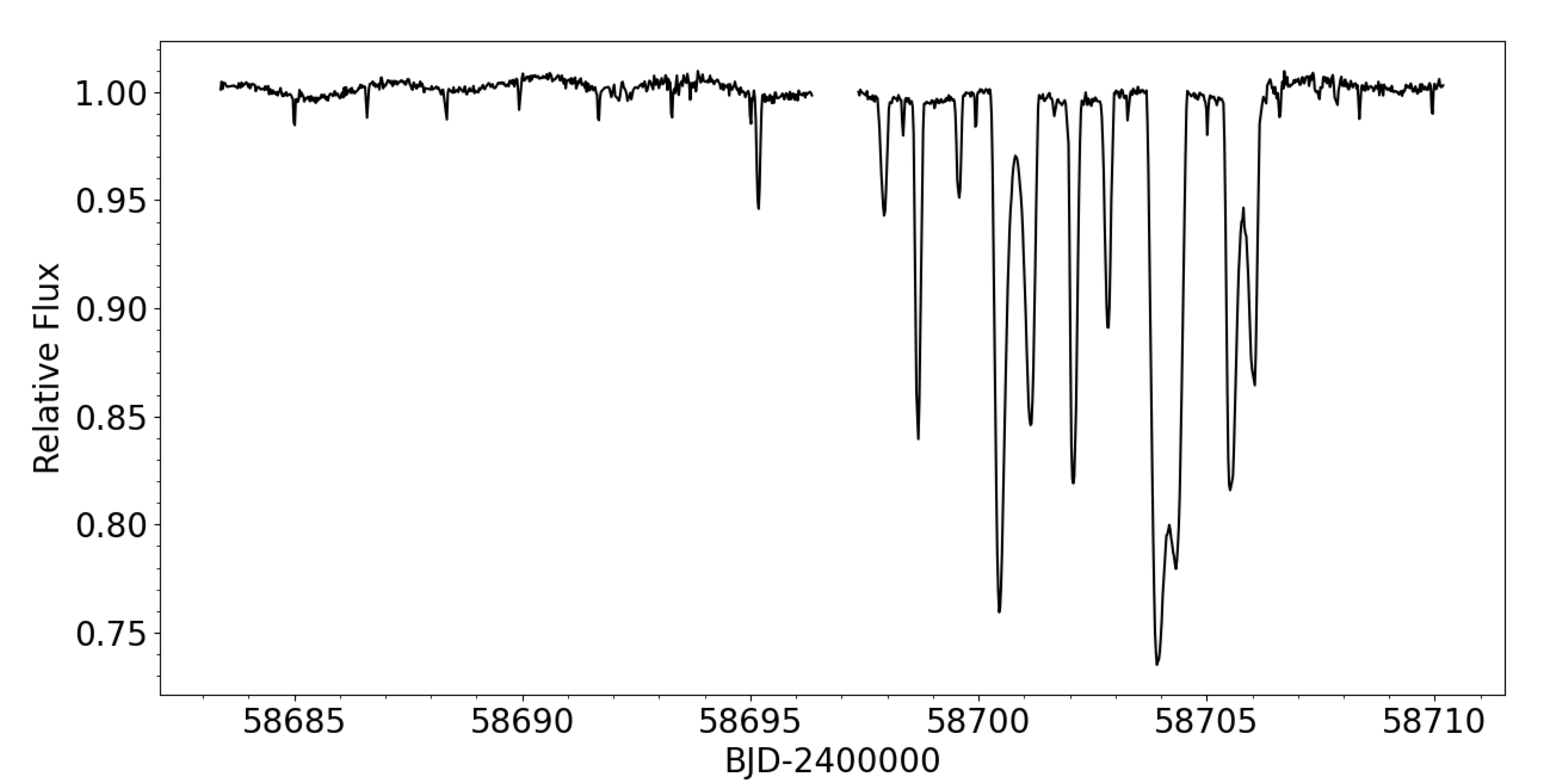}
    \includegraphics[width=1.0\linewidth]{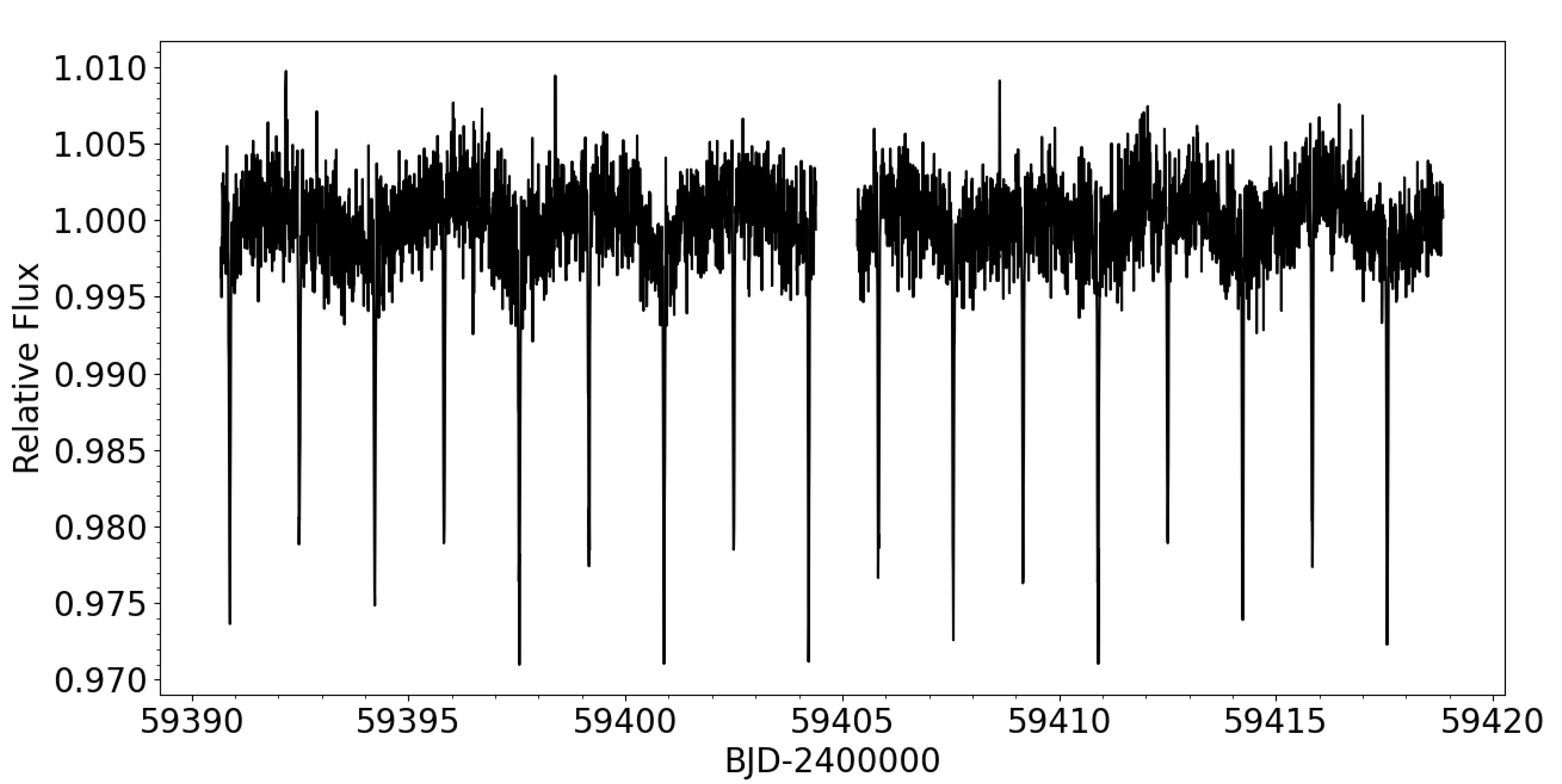}
    \includegraphics[width=1.0\linewidth]{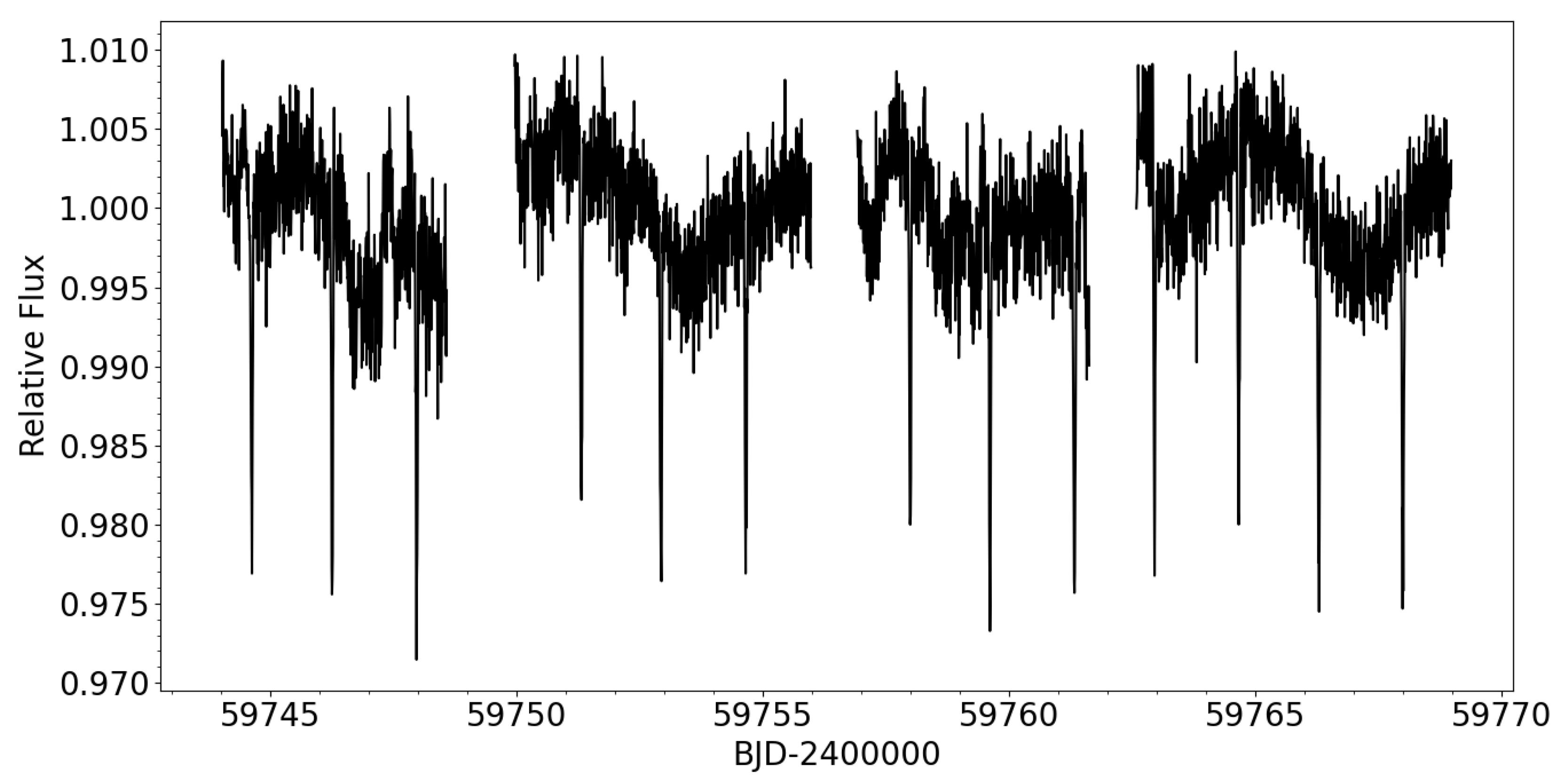}
    \caption{{\em TESS} light curves of TIC 114936199 for sector 14 ({\em top panel}), sector 40 ({\em middle panel}), and sector 53 ({\em bottom panel}), constructed using \texttt{eleanor} \citep{eleanor}.}
   \label{fig:lc}
\end{figure}

TIC 114936199 was observed by {\em TESS} in sectors 14, 40, 53, and 54.  Figure \ref{fig:lc}, top panel, shows the sector 14 light curve and the third body events of interest.  The EB is clearly visible at a period of $\sim$3.33 days.  The dips from the third body event begin at BJD $\sim$2458695 and end at BJD $\sim$2458707.  We will discuss the dips during this time interval as well as the hints provided by them as to the nature of the system in Section \ref{sec:clues}.

The sector 40 light curve is shown in the bottom panel of Figure \ref{fig:lc}.  Of particular interest is the fact that the depth of the EB eclipses is $\sim$2.5\% in sector 40, but only $\sim$1\% in sector 14, revealing the precession of the orbital plane of the EB, which will be discussed further in Section \ref{sec:mcmc}. At the time of writing, the FFIs of the first orbit of sector 53 were also available through the {\em TESS} Image Calibrator \citep{2020RNAAS...4..251F}.

We initially identified TIC 114936199 through the {\em TESS} sector 14 light curve in August, 2020, as part of an effort to construct the {\em TESS} FFI light curves \citep{2022RNAAS...6..111P} and find EBs using a neural network, discussed further in \citet{2021AJ....161..162P}.  At the time, we did not pursue it further, though we began to analyze the light curve in earnest over a year later, starting in September, 2021.

Concurrent with the initiation of our analysis, we made attempts to organize ground-based photometric follow up observations of this target. Thus, the system was observed on one night with the 80-cm RC telescope of Gothard Astrophysical Observatory, Szombathely, Hungary, in the same manner described in \citet{2022MNRAS.510.1352B}, and on two nights with the 0.65-m Mayer telescope at the Ond\v{r}ejov observatory. This reflecting telescope is operated jointly by the Astronomical Institute of ASCR and the Astronomical Institute of the Charles University of Prague, Czech Republic. The instrument is equipped with a Moravian Instruments G2-3200 CCD camera (with a Kodak KAF-3200ME sensor and standard BVRI photometric filters) mounted at the prime focus. During these three nights we were able to observe two additional secondary eclipses of the innermost pair, and the light curve of the descending branch of a third event was also obtained.

\section{Clues Provided By The Eclipses}
\label{sec:clues}

\begin{figure}[h]
    \centering
    \includegraphics[width=1.0\linewidth]{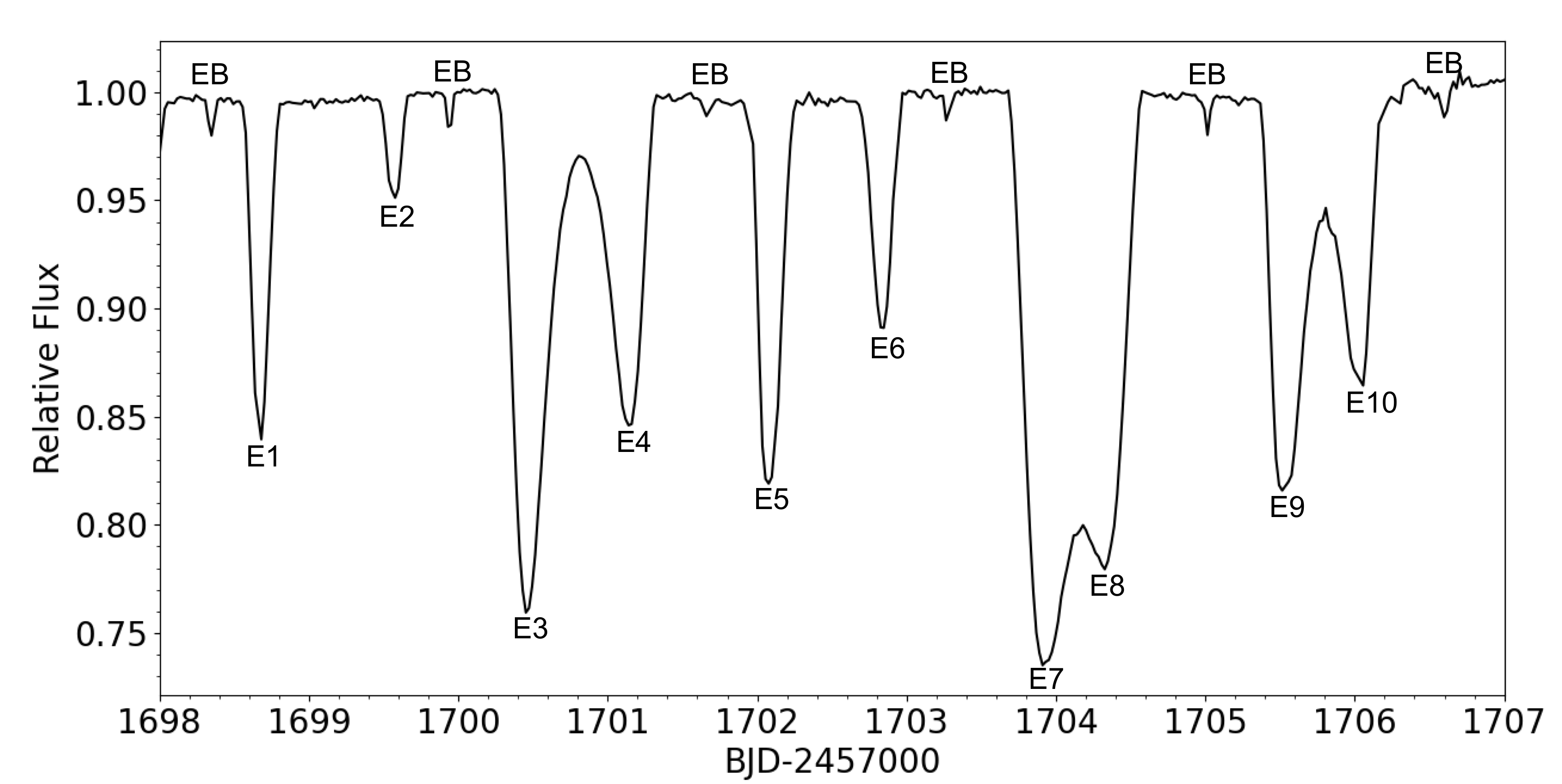}
    \caption{{\em TESS} light curve of TIC 114936199 from BJD 2458698 to 2458707, with labeled eclipses.}
   \label{fig:lc_zoom}
\end{figure}

The clearly apparent strangeness of the long transit event in {\em TESS} sector 14 was, during our initial analysis, difficult to understand as a series of eclipses among solid bodies.  Could the dips, for example, be caused by a dust cloud resulting from a disintegrating planet or asteroid, as in \citet{2021arXiv211001019P} or a seemingly random signature as in \citet{2019MNRAS.488.2455R}?  There were several clues in the light curve which led to our hypothesis that the event was indeed a series of eclipses between several stars.  In order to describe these adequately, we provide the segment of the {\em TESS} light curve for BJD 2458698-2458707 with labeled individual eclipses in Figure \ref{fig:lc_zoom}.  Those labeled ``EB'' on top are eclipses from the EB, whereas those labeled ``E'' followed by a number are from another body.  Though the actual eclipse event is several days longer than this particular analysis, we find this time period to be particularly instructive given the patterns we identify therein.

\vspace{.2cm}

The clues we identified are:

\vspace{.2cm}

(i) The dips never occur during eclipses of the EB. This suggests that the projection of the third body on the sky is passing through the EB orbit, but not through its center of mass.

\vspace{.2cm}

(ii) There are two third body dips in between each EB eclipse.  This indicates that a single star of the EB is eclipsing a third body twice, perhaps once as it moves away from the center of mass of the EB and then again as it returns.

\vspace{.2cm}

(iii)  In between the EB eclipses, five straight sets of dips occur in a clear deep-shallow pattern.  Specifically, [E1$\rightarrow$deep, E2$\rightarrow$shallow], [E3$\rightarrow$deep, E4$\rightarrow$shallow], and so on through [E9$\rightarrow$deep,E10$\rightarrow$shallow].  This pattern indicates eclipses happening with the same type of movement trajectory. 

\vspace{.2cm}

(iv)  The pattern of sets of two dips alternates such that successive sets of two are alternating deeper and comparatively shallow.  Specifically, [E1,E2]$\rightarrow$shallow, [E3,E4]$\rightarrow$deep, [E5,E6]$\rightarrow$shallow, [E7,E8]$\rightarrow$deep, [E9,E10]$\rightarrow$shallow.  Stated in a different manner, every other dip alternates such that the pattern of E1$\rightarrow$E3$\rightarrow$E5$\rightarrow$E7$\rightarrow$E9 is shallow$\rightarrow$deep$\rightarrow$shallow$\rightarrow$deep$\rightarrow$shallow.  The alternate set of dips, E2$\rightarrow$E4$\rightarrow$E6$\rightarrow$E8$\rightarrow$E10, has the same pattern: shallow$\rightarrow$deep$\rightarrow$shallow$\rightarrow$deep$\rightarrow$shallow.  This pattern indicates that there are two different stars interacting with a third body in successive sets of two dips.

\vspace{.2cm}

The clues discussed above helped us to establish a baseline concept for the model:  A third body is transitted by an EB in between its center of mass and the maximum elongation on the sky of the EB orbit such that the same star of the EB can eclipse the third body twice in the same orbit.  Additional difficulty in modeling the system is presented when we consider that the transit of the third body occurs over a $\sim$12 day duration, which requires an {\em extremely} slow outer orbit.  Alternatively, a reasonable outer orbit can be achieved if the relative motions between the binary and the eclipsed body are slowed on the sky by a fourth star, either in the form of a 2+1+1 quadruple or a 2+2 quadruple.  Spectroscopic observations presented in Section \ref{sec:spec} show that it is definitively the former, but in the next section we discuss our initial modeling approach, prior to obtaining spectral data.

\section{Orbital approximations}
\label{sec:orbits}

Having established logically that the third-body events must be due to an EB occulting a third star, the first hypothesis to consider is that this is a triple star system.  Some simple estimates indicated to us early on that this was not plausible.  We repeat that argument here because it is instructive.  Consider a triple star system with three equal masses of 0.5 M$_\odot$ (we take this rough estimate from the {\em TESS} Input Catalog \citep{2019AJ....158..138S} report of a $T_{\rm eff}$ for this system of $\sim$3900 K).  The orbital size of the 3.3-d EB is then 9.3\,R$_\odot$, and depends on only the 1/3 power of the assumed masses. If the third-body events last for 12 days, then the speed of the EB relative to the third star on the sky must average about 7 km s$^{-1}$.  From this we can infer that the period of the outer orbit, if it is assumed to be circular, is 155 years with a semimajor axis of 36 AU.  From the fact that there are several third-body eclipses for every EB orbit, we can infer that the system is nearly coplanar.  Also, since the eclipse probability in that case is $\mathcal{P} \simeq 2 R/a$, where $R$ is the stellar radius and $a$ is the outer semimajor axis, we find a rough eclipse probability for the triple-star scenario of $\sim10^{-4}$.  Finally, when we factor in the probability of catching the third-body event during a 25-day {\it TESS} observation, of $\sim25/(155 \cdot 365) \simeq 4 \times 10^{-4}$, the overall probability of observing such an event is $\sim4 \times 10^{-8}$.  This seems implausibly low even though we have examined some $10^7$ stars visually with {\it TESS} and {\em Kepler} \citep{2022PASP..134g4401K}, especially when considering that only a fraction of the stars observed are actually triples.
 
\begin{figure}[h]
    \centering
    \includegraphics[width=0.7\linewidth]{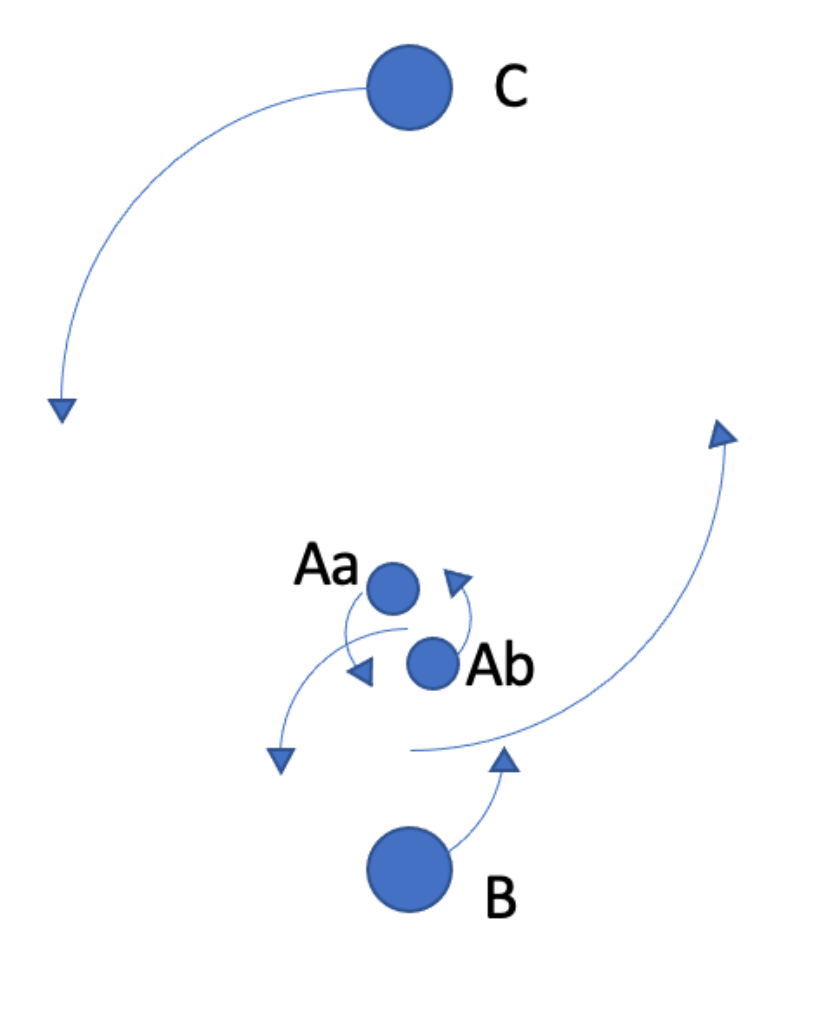}
    \caption{Schematic of a 2+1+1 quadruple, showing how star B would slow down binary A relative to star C on the sky.  Both the center of mass of the A binary and star C would, for a small fraction of the orbit of the AB binary, be moving in approximately the same direction, and with comparable speed.}
   \label{fig:abc_schematic}
\end{figure}

Instead, we then considered the possibility of a quadruple system with either a 2+2 or a 2+1+1 configuration.  In the nomenclature we will adopt in this paper, the EB consists of stars Aa and Ab, the star that is occulted during the third-body events is star C, and star B is introduced for the purpose of slowing the motion of the EB on the sky relative to star C, as demonstrated in Figure \ref{fig:abc_schematic}.  As we will show in this work, the correct interpretation of the observations is unique to the 2+1+1 configuration.  We show how we initially estimated the inner triple period as well as the outer period based on some simple arguments using Kepler's third law.  The arguments are much the same whether the configuration is 2+2 or 2+1+1.  

For purposes of estimating the orbital periods, we assume that the stellar masses are $M_{\rm Aa}$ and $M_{\rm Ab}$ (with $M_{\rm Aa}+M_{\rm Ab} \equiv M_{\rm A}$), $M_{\rm B}$, and  $M_{\rm C}$, all orbits are circular, and the triple and quadruple periods are $P_{\rm trip}$ and $P_{\rm quad}$, respectively.  The speed of C in its outer orbit relative to the center of mass (CM) of the AB triple is $v_{\rm out} = \Omega_{\rm quad} a_{\rm quad}$.  By comparison, the speed of A within the AB triple is $v_{\rm in,A} = (M_{\rm B}/M_{\rm trip}) \Omega_{\rm trip} a_{\rm trip}$.  If, during the course of the third-body event, the motion of A on the sky within the AB triple is opposite to the relative motion of AB with respect to C, then the duration of the third body eclipses can be considerably extended.  To maximize this effect, we simply set $v_{\rm in,A} \approx v_{\rm out}$, though the two only have to match, on average, during the third body eclipses, to within $\sim$7 km s$^{-1}$.  This amounts to 
\begin{equation}
 \Omega_{\rm quad} \,a_{\rm quad} \simeq (M_{\rm B}/M_{\rm trip})  \Omega_{\rm trip} a_{\rm trip}
\end{equation}
which can be simplified to
\begin{equation}
\frac{P_{\rm quad}}{P_{\rm trip}} \simeq \frac{M_{\rm trip}^2 M_{\rm quad}}{M_{\rm B}^3 }
\end{equation}
If we now illustrate this result for an all equal-mass system, then the right side of this expression has a value of 36.  Finally, for maximum dynamical stability, we might guess that $P_{\rm quad}/P_{\rm trip} \approx P_{\rm trip}/P_{\rm EB}$.  These latter two facts are sufficient for us to conclude that the likely periods are: $\sim$ 3.3 days, 120 days, and 4300 days, respectively.  In fact, as we shall see, the actual latter two periods are roughly half these values, rendering the inner triple a rather `tight' configuration.  However, this crude initial estimate of the periods did enable us to more efficiently begin to explore the parameter space while searching for a solution.

\section{Spectroscopy}
\label{sec:spec}
TIC 114936199 was observed on 11 separate occasions with the Tillinghast Reflector Echelle Spectrograph (TRES; \citealt{2007RMxAC..28..129S}; \citealt{Furesz:2008}), which is attached to the 1.5 m telescope at the Fred Lawrence Whipple Observatory on Mount Hopkins, Arizona.  The  wavelength range 3900--9100~\AA\ is covered in 51 orders at a resolving power of 44,000. 
Signal-to-noise ratios at a wavelength of 5200~\AA\ range from about 12 to 17 per resolution element
of 6.8~km~s$^{-1}$.

Initially, only a single set of relatively sharp lines was identified in the spectra at this wavelength. Preliminary radial velocities (RVs) obtained by cross-correlation against an appropriate synthetic spectrum ($T_{\rm eff} = 4000$~K) indicated little to no motion within the scatter, except for the first measurement obtained six months earlier than the others, which was slightly more negative. We now identify this object as star C. Preliminary modeling prior to obtaining RVs, discussed further in Section \ref{sec:system_model}, suggested a period of roughly 50 days for the object perturbing the motion of the eclipsing binary (i.e., star B), as well as a mass similar to or slightly smaller than star C. With this information, we focused on a redder spectral order between about 7060~\AA\ and 7160~\AA\ containing strong TiO features common in late-type objects. This facilitates their detection and the measurement of radial velocities. We then used the two-dimensional cross-correlation algorithm TODCOR \citep{1994ApJ...420..806Z} with an observed spectrum of Barnard's star (GJ~699) as the template, and were able to identify the lines of star B. Its RVs show obvious variations with a period near 51 days in an orbit of modest eccentricity ($e \approx 0.2$).

The use of TODCOR significantly improved the velocities of star C, revealing a clear but very slow upward trend, indicative of a very long period for that star. A similar but opposite trend was seen in the residuals of star B from its 51-day orbit, strongly supporting the argument for a 2+1+1 configuration in this quadruple system. The radial velocities we measured for both objects are given in Table~\ref{tab:rv}. The components of the inner eclipsing pair are much fainter and are not detected in our spectra.

\begin{table}
\center
\caption{Radial Velocities}
\label{tab:rv}
\begin{tabular}{l   c c } 
\hline
\hline
BJD        &     B    &   C   \\
$-2400000$ & \multicolumn{2}{c}{ (km~s$^{-1}$) } \\ 
\hline
59490.6117  &  -52.72  & -56.78  \\
59664.9896  &  -44.89  & -50.86  \\
59676.9835  &  -16.04  & -50.73   \\
59691.9138  &  -44.57  & -50.06   \\
59695.8780  &  -69.86  & -50.64   \\
59703.8786  &  -79.97  & -49.92   \\
59714.8344  &  -49.17  & -49.20   \\
59724.9142  &  -23.14  & -48.85   \\
59734.9113  &  -14.84  & -48.77   \\
59742.8246  &  -47.07  & -48.55   \\
59749.9384  &  -83.99  & -48.69   \\
\hline
\end{tabular} 

{Notes. The orbital solution fit to these measurements is shown in Figure \ref{fig:rvfit}.  The specification of the measurement as belonging to B or C is determined by this fit.}

\end{table}

\begin{figure}[h]
    \centering
    \includegraphics[width=1.0\linewidth]{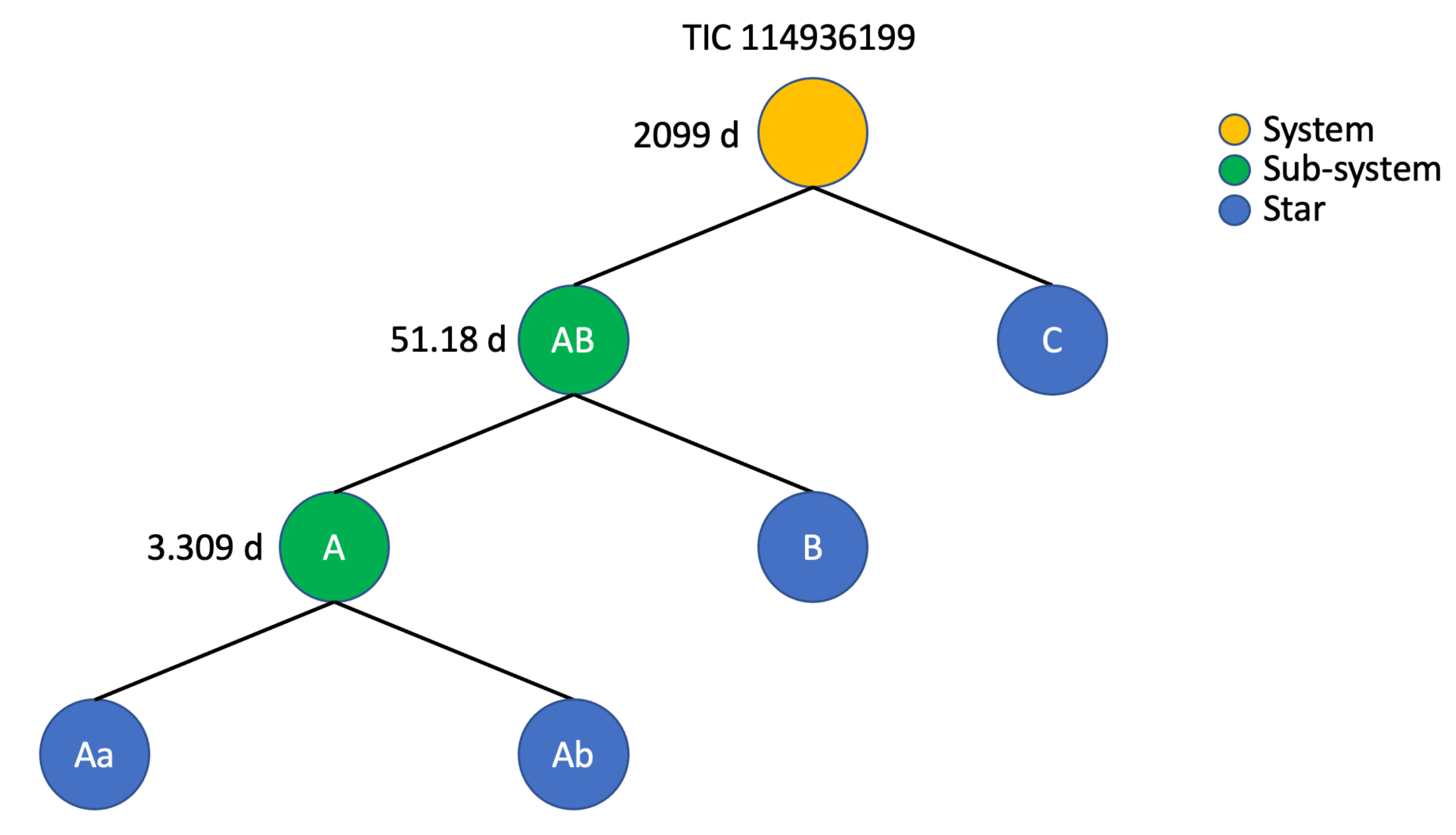}
    \caption{Schematic of TIC 114936199, with periods of each orbit.}
   \label{fig:schematic}
\end{figure} 

\begin{figure*}
    \centering
    \includegraphics[width=0.70\linewidth]{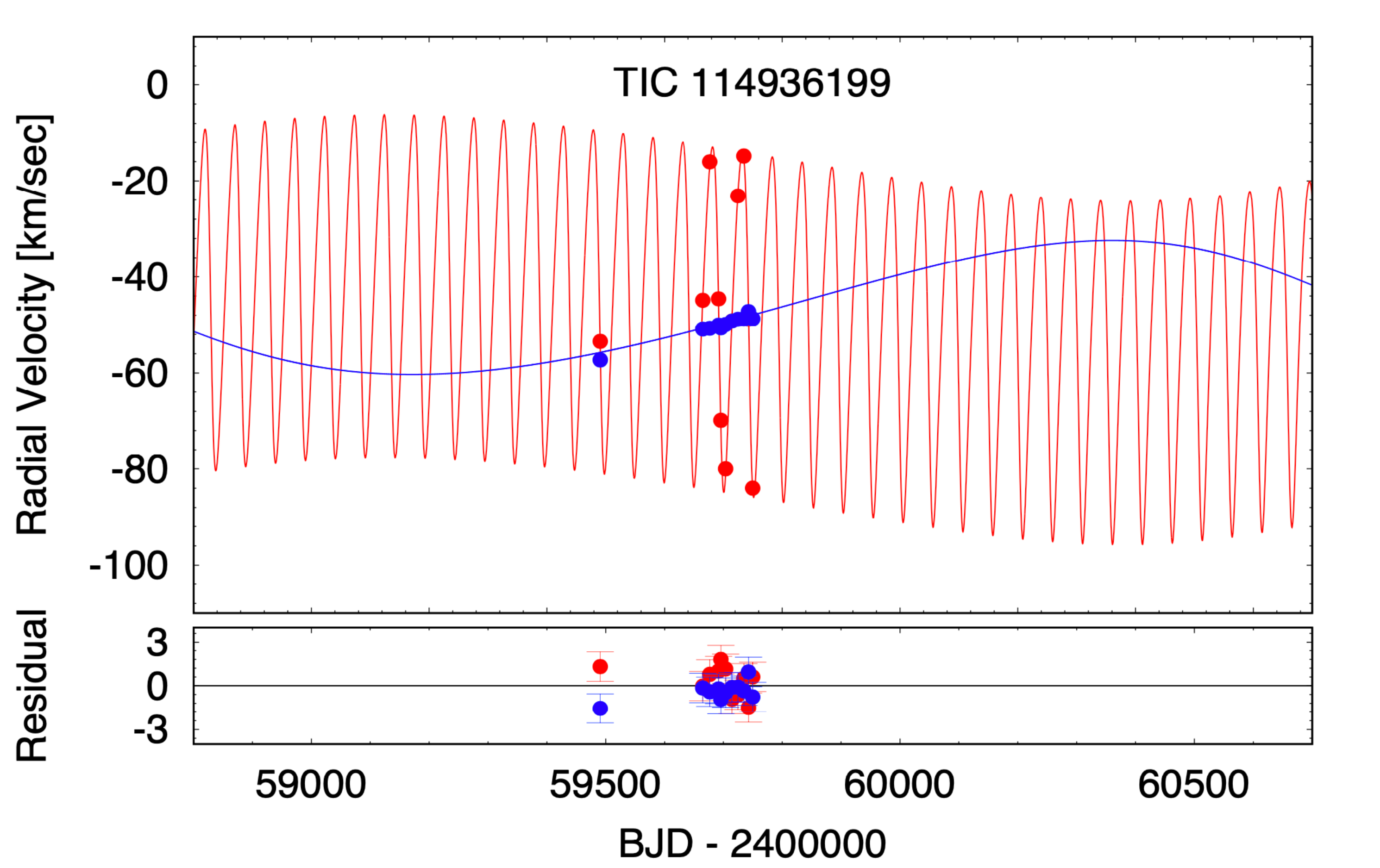}
    \caption{The fit of our model (see Section \ref{sec:mcmc}) to the TRES RVs shows two orbits: one at 51.18 days and the other 2099 days.  These are taken to be the orbits of the inner triple, AB, and the quadruple, ABC, respectively.  Our model RVs are shown for star B in red and star C in blue.}
   \label{fig:rvfit}
\end{figure*}

The RVs were critical to finding the solution to TIC 114936199 in both confirming the configuration of the system and allowing for resolution of the orbits of the inner triple and the quadruple.  From here on we will refer to the stars in the 2+1+1 configuration as Aa, Ab, B, and C, in the configuration ((Aa+Ab)+B)+C.  The EB is A, the triple AB, and the quadruple ABC, with a schematic of the system shown for reference in Figure \ref{fig:schematic}. The fit to the RV data points is shown in Figure \ref{fig:rvfit}.  We know that the RVs do not belong to either of the stars in A, as the rate of change is clearly not compatible with a $\sim$3.33 day binary.

\section{Modeling Process}
\label{sec:system_model}

In order to model the system, we wrote a spectro-photodynamical code using the \texttt{Rebound} Python library \citep{2012A&A...537A.128R} to simulate the stellar orbits combined with the \texttt{Shapely} Python library \citep{shapely} to calculate visible area of the stars.  We also incorporated MIST tracks \citep{2016ApJS..222....8D,2016ApJ...823..102C} to model the stellar evolution.

The difficulty of finding a solution to the system model required an initial departure from conventional MCMC modeling.  The number of free parameters in our 2+1+1 quadruple star system (23; the four stellar masses and their assumed coeval age, the orbital parameters of three distinct orbits, and the system gamma velocity) and their extremely wide bounds, given our limited knowledge of the system, were beyond the capability of MCMC to solve. To this end, we employed heuristic optimization algorithms.  

Initially, we used the Particle Swarm Optimization (PSO) algorithm \citep{488968} via the PySwarms Python library \citep{pyswarmsJOSS2018}.  Briefly, PSO randomly chooses multiple sets of model parameters within given bounds, each set of parameters comprising a group called a ``particle'', iteratively moving the parameters of each set toward its own best solution as well as the best solution of the group (or ``swarm'') in search of a minimum. Though this algorithm was sufficient for basic modeling, we found that it inadequately sampled the parameter space and had a tendency to converge on spurious local minima.  

We made substantial progress with PSO, but ultimately switched our approach to using the Differential Evolution (DE) algorithm \citep{Storn1997DifferentialE}, specifically the SciPy Python implementation \citep{scipy}, which we found offered a more thorough exploration of the parameter space.  Briefly, DE maintains a candidate set of solutions known as the ``population'', which is perturbed at every iteration by introduced random ``mutations'' which improve the diversity of the population.  We found DE to be quite powerful at searching our rather vast parameter space.  

This is not to say that any given use of DE would solve the system.  In fact, it did not.  Heuristics, in general, make no guarantee of convergence, and DE is no exception.  To this end, we employed the {\em Discover} supercomputer at the NASA Center for Climate Simulation (NCCS)\footnote{\url{https://www.nccs.nasa.gov/systems/discover}}.  Using {\em Discover}, we ran multiple parallelized instances of the same problem in a loop, saving the solution and the minimum $\chi^{2}$ achieved with each iteration.  Parallelization was critical to the time required to solve this problem in order to use, for PSO, more particles, and, for DE, a large population. For DE specifically, we used a population size of 36, with each sample running on a separate core of a node on {\em Discover}.  A standard computational run for us consisted of 12 nodes, each operating 36 cores in parallel, for three days, totaling over $30,000$ CPU-hours for any given model run.  We attempted on the order of hundreds of these model runs with various parameter bounds and modifications to the composition of $\chi^{2}$, though we terminated many of them early after poor results.

Our modeling process was highly iterative, over the course of several months of effort.  In order to first confirm our hypothesis about the nature of the eclipses derived from the clues noted in Section \ref{sec:clues}, our initial modeling consisted of an EB with a stationary center of mass and a third star passing behind it in a non-physical motion.  We used the PSO algorithm to minimize the error in this basic model.  After establishing that our hypothesis was largely sufficient to create the remarkable eclipse pattern from the {\em TESS} sector 14 light curve, we proceeded to develop the physical model with approximate stable orbits theorized in Section \ref{sec:orbits}.  We first attempted the physical model using PSO, but ultimately switched to DE, with which we were quite pleased.

This model fit was set up to minimize $\chi^{2}$, which included the {\em TESS} light curves from sectors 14 and 40, the RV data points from TRES, and the system $T_{\rm eff}$ from the TIC as compared to the $T_{\rm eff}$ of star C as a proxy for the SED with lower computational cost.  We were able, using DE in parallel on the {\em Discover} system, to find an approximate solution to the system.  We note here that the periods of AB and ABC as found by the DE model, were substantially shorter than those estimated as shown in Section \ref{sec:orbits}.  The bounds given to the DE solver for these orbits were quite broad, and the estimates provided worthwhile guides.

It is important to note that we used DE simply as a starting point for MCMC analysis.  We needed to identify a parameter space from which MCMC could find a viable solution, which we indeed accomplished.  DE found the coarse approximate solution, while MCMC provided the refined solution.  MCMC would not have been capable of finding approximate parameters in any reasonable amount of time, while DE cannot provide a probability distribution for the parameters of the final solution.  The pairing of these two methods was absolutely critical to solving this system.  

\begin{table*}
\caption{Binary Mid Eclipse Times for TIC~114936199}
 \label{tab:T114936199ToM}
\begin{tabular}{lrllrllrl}
\hline
BJD & Cycle  & std. dev. & BJD & Cycle  & std. dev. & BJD & Cycle  & std. dev. \\ 
$-2\,400\,000$ & no. &   \multicolumn{1}{c}{$(d)$} & $-2\,400\,000$ & no. &   \multicolumn{1}{c}{$(d)$} & $-2\,400\,000$ & no. &   \multicolumn{1}{c}{$(d)$} \\ 
\hline
58688.343254 &    0.5 & 0.001410 & 59399.160610 &  213.5 & 0.000457 & 59744.636890 &  317.0 & 0.000283 \\
58691.677763 &    1.5 & 0.001312 & 59400.891995 &  214.0 & 0.000423 & 59746.268401 &  317.5 & 0.000433 \\
58693.277155 &    2.0 & 0.002179 & 59402.498408 &  214.5 & 0.000429 & 59747.981496 &  318.0 & 0.000265 \\
58695.008248 &    2.5 & 0.001552 & 59404.227150 &  215.0 & 0.000351 & 59749.606306 &  318.5 & 0.000455 \\ 
58698.344784 &    3.5 & 0.001491 & 59405.836317 &  215.5 & 0.000354 & 59751.326061 &  319.0 & 0.000204 \\ 
58699.939543 &    4.0 & 0.001845 & 59407.561919 &  216.0 & 0.000443 & 59752.941843 &  319.5 & 0.000215 \\ 
58701.679696 &    4.5 & 0.001053 & 59409.171049 &  216.5 & 0.000354 & 59754.663598 &  320.0 & 0.000493 \\ 
58703.276991 &    5.0 & 0.002179 & 59410.897680 &  217.0 & 0.000352 & 59757.999171 &  321.0 & 0.000699 \\ 
58705.014211 &    5.5 & 0.000835 & 59412.506270 &  217.5 & 0.000272 & 59759.617965 &  321.5 & 0.000401 \\  
58706.618479 &    6.0 & 0.003207 & 59414.233527 &  218.0 & 0.000395 & 59761.333652 &  322.0 & 0.000306 \\
59390.878459 &  211.0 & 0.000320 & 59415.839228 &  218.5 & 0.000332 & 59764.666965 &  323.0 & 0.000384 \\
59392.483341 &  211.5 & 0.000496 & 59417.568810 &  219.0 & 0.000237 & 59766.291773 &  323.5 & 0.000224 \\
59394.222050 &  212.0 & 0.000361 & 59709.547823 &  306.5 & 0.000211 & 59768.000692 &  324.0 & 0.000500 \\
59395.820709 &  212.5 & 0.000551 & 59719.559041 &  309.5 & 0.000122 & & & \\
\hline
\end{tabular}

{\it Notes.} Integer and half-integer cycle numbers refer to primary and secondary eclipses, respectively. Eclipses with cycle numbers $0.5-6.0$, $211.0-219.0$ and $317.0-320.0$ were observed by the \textit{TESS} spacecraft during sectors 14, 40 and 53, respectively. The two secondary eclipse times with cycle numbers $306.5$ and $309.5$ were determined from ground-based follow up observations. The last six eclipses (cycle nos. $321.0-324.0$) became available after the completion of our analysis and, hence, were not used in our photodynamical modeling process.
\end{table*}

\section{MCMC Model}
\label{sec:mcmc}

After finding an approximate solution with DE and the \texttt{Rebound} model, we proceeded to use \texttt{Lightcurvefactory} \citep{2019MNRAS.483.1934B,2020MNRAS.493.5005B} to produce an MCMC model of the system.  This model simultaneously fit light curves (including {\em TESS} sectors 14, 40, 53, and three additional ground-based, follow up regular eclipse observations of the innermost pair), the ETV curve (determined from the midpoint times of the regular eclipses observed with {\em TESS} and ground-based telescopes, see Table~\ref{tab:T114936199ToM}), the TRES RV measurements, and the net SED of the quadruple system. Moreover, the model also applies simultaneously built-in \texttt{PARSEC} three dimensional (mass; age; metallicity) isochrone grids \citep{2012MNRAS.427..127B} to constrain the astrophysical parameters and evolutionary states of the stars under investigation. In each MCMC run we had 24 freely adjusted parameters. These parameters are mainly dynamical and orbital as well as the masses of the two outer stars, B and C ($m_\mathrm{B,C}$); the mass ratios of the innermost and middle orbits ($q_{1,2}$); the inclinations ($i$) and eccentricity vectors ($e\sin\omega$, $e\cos\omega$) of all the three orbits; the longitudes of the nodes of the middle and outer orbits ($\Omega_{2,3}$) on the tangential plane of the sky, relative to the ascending node of the innermost orbit (i.e., this latter was set to $\Omega_1=0$); and the times of the conjunctions of the stars on the middle and outer orbits ($\mathcal{T}_{2,3}^\mathrm{inf/sup}$) which occurred during the outer eclipsing activity in Sector 14.\footnote{As we will discuss soon, these latter parameters were found to be the most critical ones. Note, it was found already from the former DE solution (see Sect.~\ref{sec:system_model} above) that in case of the middle orbit, star B was in an inferior conjunction ($\mathcal{T}_2^\mathrm{inf}$) with the center of mass of the inner pair, while for the outmost orbit, star C was in a superior conjunction ($\mathcal{T}_3^\mathrm{sup}$) with the center of mass of the inner triple subsystem.}  

Additionally, when we use the \texttt{PARSEC} grids, the metallicity [M/H] and (logarithmic) age of the quadruple system (assuming strictly coeval stars) were also adjusted, while for the SED analysis the interstellar reddening $E(B-V)$ was also introduced as a freely adjustable trial parameter. Finally, we allowed the amounts of the extra light (i.e., contaminating flux) to be adjusted in both kinds of light curves we used, i.e., in \textit{TESS}- and Cousins $R_\mathrm{C}$-bands. Several additional parameters were internally constrained and calculated by the software package during each MCMC trial step, as follows. The fundamental parameters of the four stars, necessary for the light curve emulation (i.e., radii, effective temperatures) were iterated with the use of the \texttt{PARSEC} grids \citep[see][for the detailed description of the process]{2020MNRAS.493.5005B}. Furthermore, the period of the innermost binary ($P_1$) and the mid-time of the very first regular eclipse (a secondary minimum, i.e., a superior conjunction of star Ab -- $\mathcal{T}_1^\mathrm{sup}$) were constrained with the use of the ETV curve in the manner described in \citet{2019MNRAS.483.1934B}. Moreover, the systemic radial velocity of the quadruple system, and its distance from the Earth were calculated at each iteration step a posteriori, minimizing the values of $\chi^2_\mathrm{RV}$ and $\chi^2_\mathrm{SED}$, respectively. 

Before the initialization of the MCMC runs, we made some further preparatory steps on the light curves. We detrended the \textit{TESS} light curves (with the exception of that part where the extra eclipses occurred) with the use of 2nd-8th order smoothing polynomials to the out-of-regular-eclipse sections of the light curves. In such a manner, we removed any systematics and produced light curves with flat out-of-eclipse sections. In the process, besides instrumental effects, we might have also removed some real physical light curve distortions, e.g., arising from some chromospheric activity or a low-amplitude ellipsoidal light variation. However, since we did not intend to model such extra effects, we found these flat light curves to be more appropriate for our purposes. Moreover, in order to save computational time, and also increase the relative weights of the eclipses (which are the main source of information in the EB light curve), we removed the larger parts of the out-of-regular-eclipse sections during the fitting process. 

\begin{figure*}
    \centering
    \includegraphics[width=0.8\linewidth]{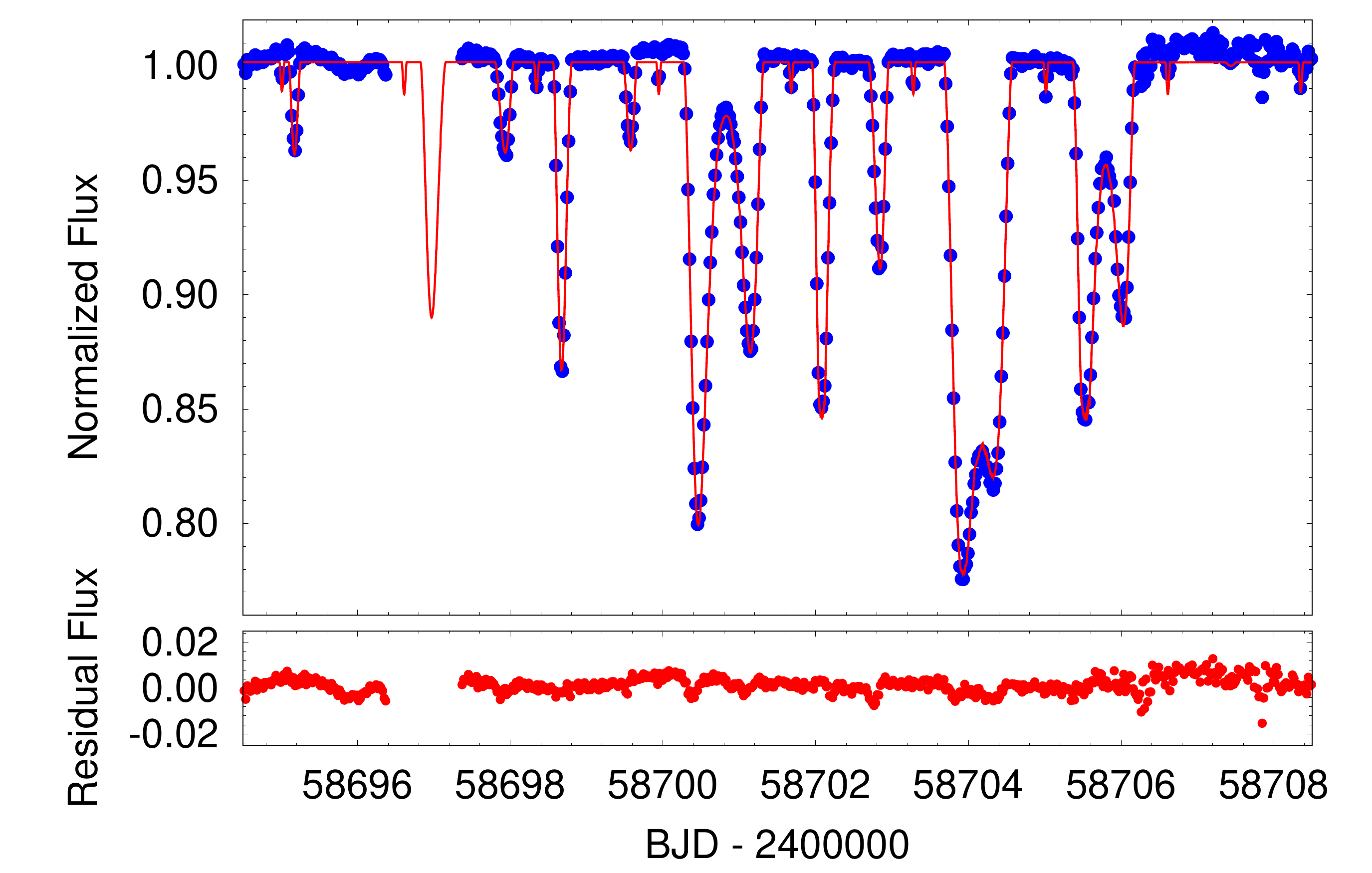}
    \caption{The {\em TESS} sector 14 data points (blue) and the model (red) curve during the third body event.  Note the predicted eclipse in the data gap. Residuals are shown in the bottom panel.}
   \label{fig:model14}
\end{figure*}

\begin{figure}
    \centering
    \includegraphics[width=0.9\linewidth]{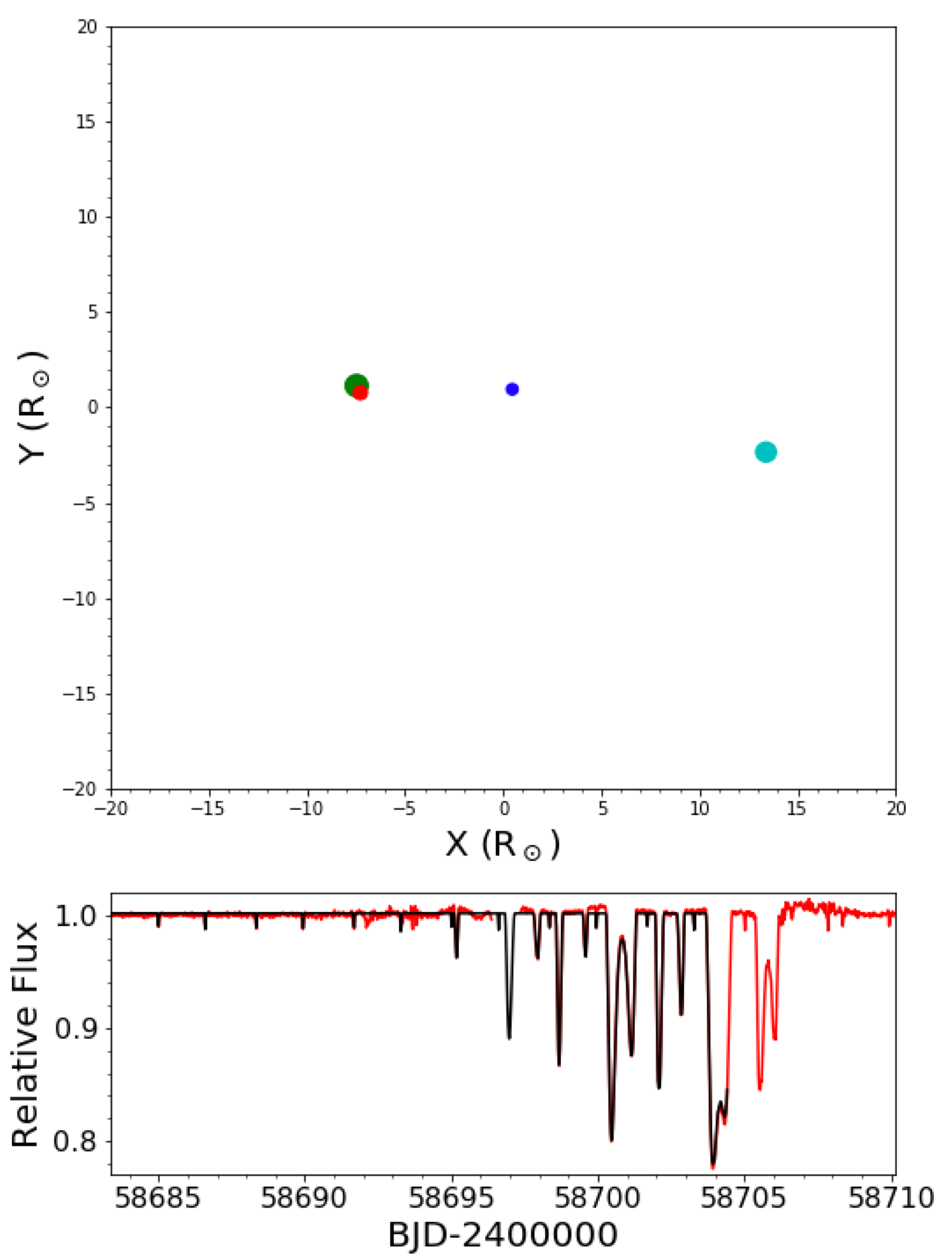}
\caption{({\em Top panel}) Positions of the stars comprising TIC 114936199 in the x-y plane as viewed from $z=-\infty$ at BJD 2458704.405.  Aa is Red, Ab is blue, B is cyan, and C is green.  ({\em Bottom panel}) The full {\em TESS} sector 14 light curve (red) plotted against the model light curve (black) concurrent with the stellar positions in the top panel. This figure is available as an animation. The animation shows the stellar positions and model light curve for the duration of {\em TESS} sector 14, with a real time duration of 50 s.}
\label{fig:animation}
\end{figure}

\begin{figure}
    \centering
    \includegraphics[width=1.0\linewidth]{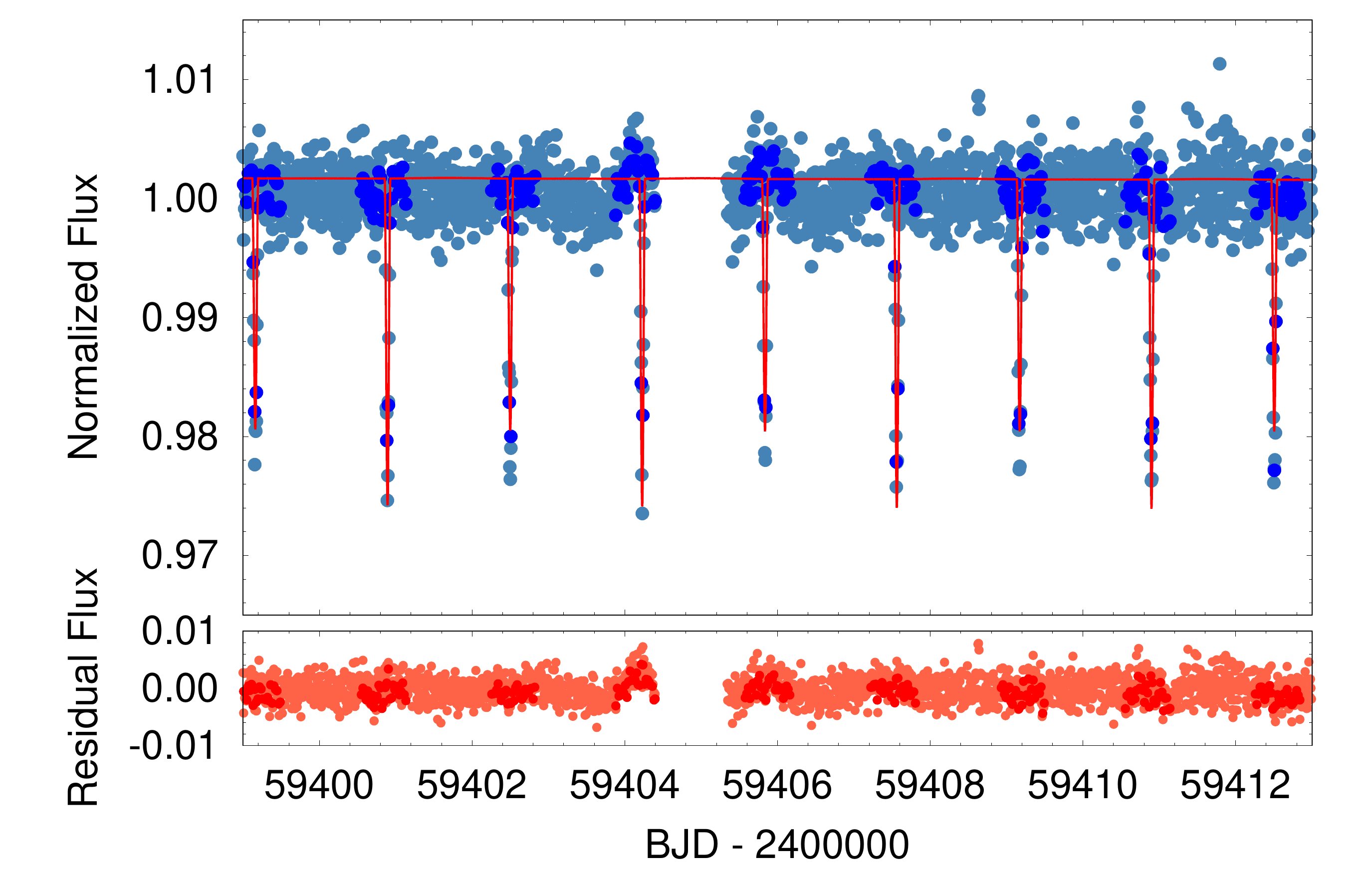}
    \caption{A section of the {\em TESS} sector 40 light curve. The pale blue points are the original (but detrended), 600-sec cadenced \texttt{Fitsh} FFI photometry points, while the dark blue around the regular eclipses denote the 1800-sec binned points used for the photodynamical MCMC modeling. Residuals are shown in the bottom panel.}
   \label{fig:model40}
\end{figure}

\begin{figure}[h]
    \centering
    \includegraphics[width=1.0\linewidth]{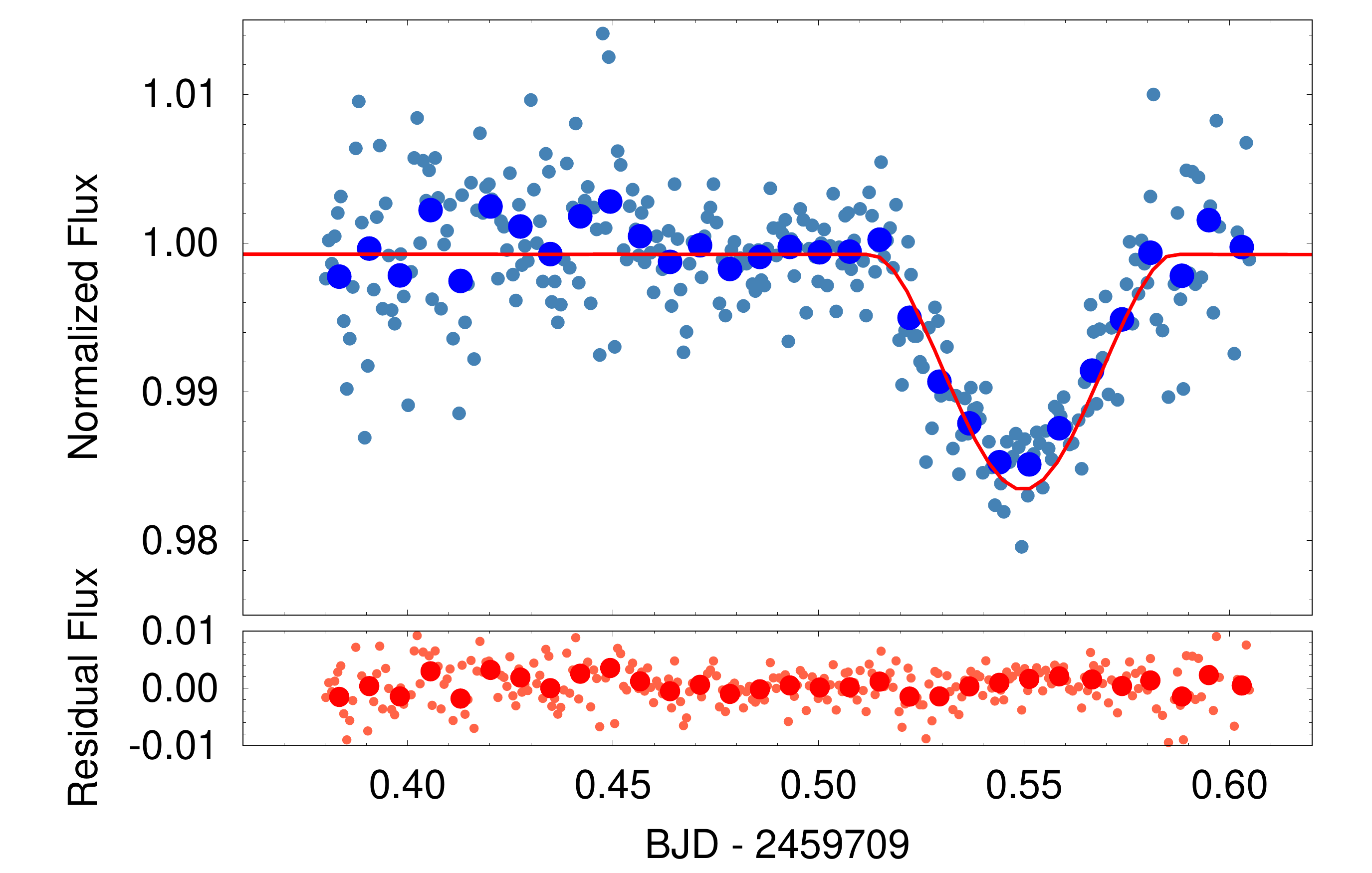}
    \caption{A secondary eclipse of the innermost binary observed by H. Ku\v{c}\'akov\'a on the night of 09/10 May 2022 with a Cousins $R_\mathrm{C}$ filter on the 0.65-m Mayer telescope of the Ond\v{r}ejov observatory, CZ. Here, the smaller, lighter blue dots represent the original 60-sec exposure time observations, while the larger, darker blue circles stand for their 600-sec bins used for the modeling. Residuals are shown in the bottom panel.}
   \label{fig:model_ground}
\end{figure}

\begin{figure*}
    \centering
    \includegraphics[width=0.7\linewidth]{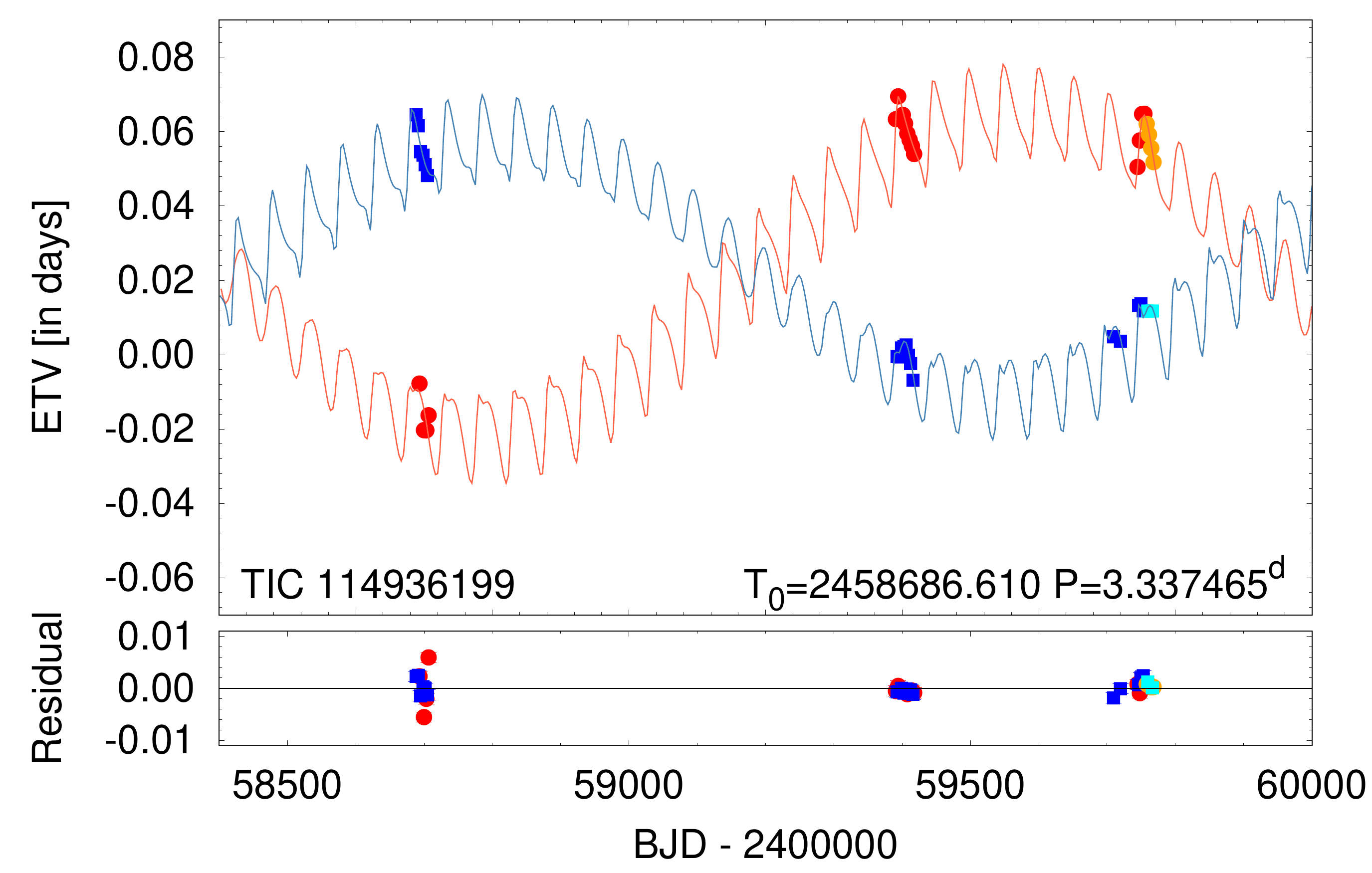}
    \caption{The model ETVs (lines) and measurements (points) for primary (red) and secondary (blue) eclipses. The longer-term sinusoidal part of the red and blue curves represents driven apsidal motion of the 3.3-day binary by star B in its 51-day orbit.  The apsidal motion period is relatively short at only $\simeq 4$ years. The more rapid sinusoids at a $\sim$51-day period represent the largely dynamical delays that result from the orbit of star B around the A binary.  These two effects alone constitute independent and dramatic proof that star B orbits the A binary. The last six points denoted with orange and cyan colors were obtained only after the completion of the photodynamical modeling process and were therefore not used for the fitting. It can be seen, however, that they align nicely with the model as predicted.}
   \label{fig:etv}
\end{figure*}

\begin{figure*}[h]
\centering
\includegraphics[width=0.48 \linewidth]{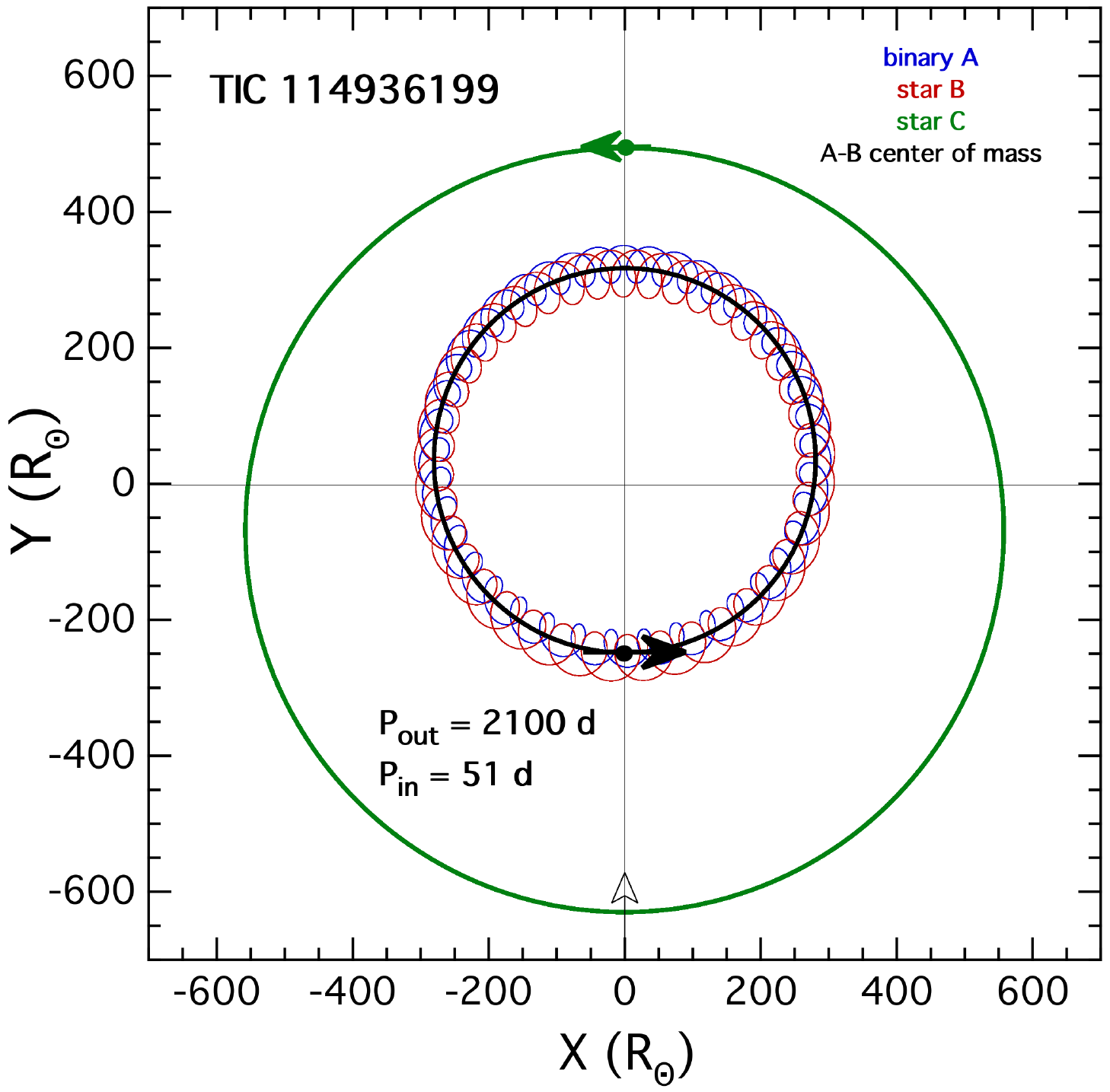} 
\includegraphics[width=0.48 \linewidth]{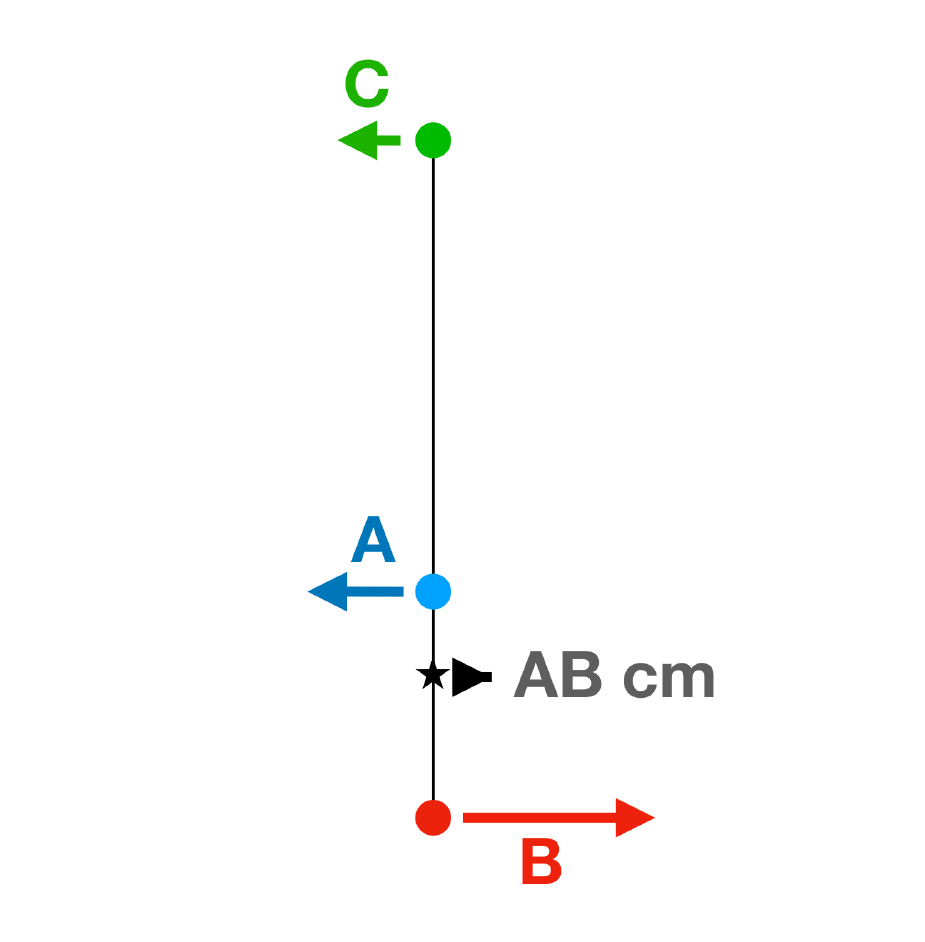}
\caption{{\it Left panel:} Orbital motions of star C (green), star B (red), center of mass of binary A (A$_{\rm cm}$; blue), and the center of mass of binary A-B (AB$_{\rm cm}$) in black, all drawn according to our final model (see Table \ref{tbl:simlightcurve}).  The observers are viewing the system from $Y = -\infty$, but the reader is looking down on the orbits from $Z = +\infty$. {\it Right panel:} Schematic showing how the velocities of the stars conspire to slow the relative motion of star C and A$_{\rm cm}$ on the sky during {\em TESS} sector 14, thereby making the third body eclipses last for $\sim$12 days. While the motion of AB$_{\rm cm}$ is opposite to that of C (thereby causing a high relative speed), the orbital motion of A$_{\rm cm}$ within the A-B binary is opposite that of AB$_{\rm cm}$ only at this particular phase of the A-B binary, and is therefore in the same direction as that of star C, thereby slowing the relative motion.  Note that the velocities are scaled approximately correctly, while the vertical separation of the stars is not at all to scale.}
\label{fig:orbits} 
\end{figure*}

The results of our spectro-photodynamical analysis are reported in Table~\ref{tbl:simlightcurve}.  There we tabulate the median values of the main system parameters together with their statistical $1\sigma$ uncertainties. In Fig.~\ref{fig:model14} we show the photodynamical model fit to the third body events (seen in Sector 14) superposed on the data points, with an animation of the system for the duration of Sector 14 shown in the online paper in Figure \ref{fig:animation}, whereas a still frame of this animation is in the static version of the paper. All of the extra (i.e., outer) eclipses are beautifully accounted for. Likewise, in Fig.~\ref{fig:model40} we show the photodynamical model for a segment of the regular EB eclipses observed in Sector 40. Figure \ref{fig:model_ground} shows the fit to one of the regular EB eclipses observed from the ground. Finally, the ETV curve obtained from the three {\em TESS} sectors and the ground based photometry is shown in Fig.~\ref{fig:etv} along with a model fit to ETV points.  Also, to help the reader better visualize the orbits of the system and the nature of the relative motions during the outer eclipse event in {\em TESS} sector 14, we provide an orbital diagram and schematic in Figure \ref{fig:orbits}.

\begin{table*}
\centering
\caption{Median values of the parameters from the simultaneous light curve, radial velocity, ETV, joint SED and \texttt{PARSEC} 
evolutionary track solution from {\sc Lightcurvefactory.}}
\begin{tabular}{lcccc}
\hline
\multicolumn{5}{c}{Orbital elements$^a$} \\
\hline
   & \multicolumn{4}{c}{subsystem}  \\
   & \multicolumn{2}{c}{Aa--Ab} & A--B & AB--C \\
  \hline
$P$ [days]                       & \multicolumn{2}{c}{$3.30889_{-0.00015}^{+0.00018}$} & $51.1750_{-0.0051}^{+0.0052}$       & $2099_{-16}^{+22}$ \\
semimajor axis  [$R_\odot$]      & \multicolumn{2}{c}{$8.23_{-0.09}^{+0.10}$}          & $62.0_{-0.7}^{+0.7}$                & $849_{-13}^{+5}$ \\  
$i$ [deg]                        & \multicolumn{2}{c}{$86.90_{-0.08}^{+0.10}$}         & $87.47_{-0.05}^{+0.04}$             & $89.857_{-0.002}^{+0.001}$  \\
$e$                              & \multicolumn{2}{c}{$0.0416_{-0.0002}^{+0.0002}$}    & $0.2196_{-0.0020}^{+0.0024}$        & $0.1211_{-0.0047}^{+0.0066}$ \\  
$\omega$ [deg]                   & \multicolumn{2}{c}{$141.0_{-0.3}^{+0.2}$}           & $86.9_{-0.8}^{+0.8}$                & $65.3_{-7.4}^{+3.4}$ \\
${\mathcal{T}^\mathrm{inf/sup}}^b$  [BJD] & \multicolumn{2}{c}{${2\,458\,684.9960_{-0.0004}^{+0.0004}}^*$}         & $2\,458\,701.632_{-0.033}^{+0.047}$               & ${2\,458\,700.464_{-0.056}^{+0.044}}^*$ \\
$\tau^c$ [BJD]                   & \multicolumn{2}{c}{$2\,458\,682.122_{-0.002}^{+0.002}$} & $2\,458\,675.38_{-0.20}^{+0.17}$ & $2\,458\,588_{-33}^{+14}$\\
$\Omega$ [deg]                   & \multicolumn{2}{c}{$0.0$}                           & $-0.02_{-0.15}^{+0.15}$             & $0.90_{-0.14}^{+0.15}$ \\
$(i_\mathrm{m})_{A-...}^d$ [deg] & \multicolumn{2}{c}{$0.0$}                           & $0.59_{-0.08}^{+0.10}$              & $3.09_{-0.09}^{+0.11}$ \\
$(i_\mathrm{m})_{B-...}^d$ [deg]   & \multicolumn{2}{c}{$0.59_{-0.08}^{+0.10}$}          & $0.0$                               & $2.56_{-0.04}^{+0.05}$ \\
\hline
\multicolumn{5}{c}{RV curve related parameters} \\
\hline
mass ratio $[q=m_\mathrm{sec}/m_\mathrm{pri}]$ & \multicolumn{2}{c}{$0.7841_{-0.0017}^{+0.0018}$} & $0.7929_{-0.0027}^{+0.0031}$ & $0.503_{-0.009}^{+0.009}$ \\
$K_\mathrm{pri}$ [km\,s$^{-1}$]                & \multicolumn{2}{c}{$55.30_{-0.62}^{+0.60}$}      & $27.80_{-0.32}^{+0.32}$      & $6.86_{-0.08}^{+0.12}$ \\ 
$K_\mathrm{sec}$ [km\,s$^{-1}$]                & \multicolumn{2}{c}{$70.47_{-0.73}^{+0.88}$}      & $35.05_{-0.38}^{+0.40}$      & $13.68_{-0.21}^{+0.17}$ \\ 
$\gamma$ [km/s]                 & \multicolumn{2}{c}{$-$} & $-$ & $-46.74_{-0.12}^{+0.14}$\\  
  \hline  
\multicolumn{5}{c}{Stellar parameters} \\
\hline
   & Aa & Ab &  B & C  \\
  \hline
 \multicolumn{5}{c}{Relative quantities} \\
  \hline
fractional radius [$R/a$]               & $0.0444_{-0.0007}^{+0.0012}$ & $0.0363_{-0.0004}^{+0.0008}$  & $0.0084_{-0.0002}^{+0.0002}$ & $0.00069_{-0.00001}^{+0.00002}$ \\
temperature relative to $(T_\mathrm{eff})_\mathrm{Aa}$   & $1$ & $0.9597_{-0.0021}^{+0.0013}$  & $1.1121_{-0.0094}^{+0.0107}$ & $1.2055_{-0.0156}^{+0.0174}$ \\
fractional flux [in \textit{TESS}-band] & $0.087_{-0.005}^{+0.004}$    & $0.046_{-0.003}^{+0.002}$     & $0.326_{-0.013}^{+0.013}$    & $0.533_{-0.016}^{+0.019}$    \\
fractional flux [in $R_C$-band]& $0.059_{-0.003}^{+0.003}$    & $0.031_{-0.002}^{+0.002}$     & $0.284_{-0.021}^{+0.021}$    & $0.617_{-0.026}^{+0.026}$    \\
 \hline
 \multicolumn{5}{c}{Physical Quantities} \\
  \hline 
 $m$ [M$_\odot$]   & $0.382_{-0.012}^{+0.014}$ & $0.300_{-0.010}^{+0.010}$ & $0.540_{-0.017}^{+0.020}$ & $0.615_{-0.016}^{+0.023}$ \\
 $R^e$ [R$_\odot$] & $0.365_{-0.010}^{+0.014}$ & $0.298_{-0.005}^{+0.010}$ & $0.521_{-0.018}^{+0.019}$ & $0.584_{-0.015}^{+0.021}$ \\
 $T_\mathrm{eff}^e$ [K]& $3357_{-49}^{+19}$    & $3225_{-59}^{+17}$        & $3733_{-38}^{+42}$        & $4047_{-31}^{+40}$        \\
 $L_\mathrm{bol}^e$ [L$_\odot$] & $0.0152_{-0.0008}^{+0.0007}$ & $0.0086_{-0.0003}^{+0.0003}$ & $0.0482_{-0.0051}^{+0.0034}$ & $0.0824_{-0.0054}^{+0.0066}$ \\
 $M_\mathrm{bol}^e$ & $9.32_{-0.05}^{+0.06}$    & $9.93_{-0.04}^{+0.04}$    & $8.06_{-0.07}^{+0.12}$    & $7.48_{-0.08}^{+0.07}$  \\
 $M_V^e           $ & $11.25_{-0.09}^{+0.12}$    & $12.15_{-0.09}^{+0.18}$    & $9.41_{-0.12}^{+0.14}$    & $8.49_{-0.11}^{+0.09}$ \\
 $\log g^e$ [dex]   & $4.893_{-0.017}^{+0.010}$ & $4.959_{-0.012}^{+0.010}$ & $4.735_{-0.016}^{+0.016}$ & $4.693_{-0.014}^{+0.011}$ \\
 \hline
\multicolumn{5}{c}{Global Quantities} \\
\hline
$\log$(age) [dex] &\multicolumn{4}{c}{$8.709_{-0.067}^{+0.06}$} \\
$ [M/H]$  [dex]      &\multicolumn{4}{c}{$-0.091_{-0.025}^{+0.147}$} \\
$E(B-V)$ [mag]    &\multicolumn{4}{c}{$0.042_{-0.019}^{+0.032}$} \\
extra light $\ell_5$ [in \textit{TESS}-band] &\multicolumn{4}{c}{$0.006_{-0.005}^{+0.008}$} \\
extra light $\ell_5$ [in $R_C$-band] &\multicolumn{4}{c}{$0.007_{-0.004}^{+0.008}$} \\
$(M_V)_\mathrm{tot}^e$           &\multicolumn{4}{c}{$8.01_{-0.07}^{+0.10}$} \\
distance [pc]                &\multicolumn{4}{c}{$124_{-4}^{+4}$}  \\  
\hline
\end{tabular}
\label{tbl:simlightcurve}

{\em Notes.} (a) Instantaneous, osculating orbital elements at epoch $t_0=2\,458\,683.368980$; (b) Moment of an inferior or superior conjunction of star Ab, B and C along their innermost, middle and outmost orbits, respectively. Superior conjunctions are noted with *; (c) Time of periastron passsage; (d) Mutual (relative) inclination; (e) Interpolated from the \texttt{PARSEC} isochrones;  
\end{table*}

As one can see from Table \ref{tbl:simlightcurve}, we have found that the system is very flat. The mutual inclination of any pair of the three orbits does not exceed $\sim3^\circ$. In this regard, note that if we consider only the S14 light curve (i.e. the segment of the extra eclipses) and the RV measurements, we are also able to find an almost similarly likely flat, but {\it retrograde}, configuration by reversing the direction of the orbits of stars Aa and Ab in the innermost binary, and relabeling them (or, more strictly speaking, modifying their true anomalies by $\sim180^\circ$). On the other hand, however, once we add the S40 (Fig.~\ref{fig:model40}) and S53 \textit{TESS} observations, as well as the ground-based photometric measurements (Fig.~\ref{fig:model_ground}), this retrograde model is ruled out due to timing anomalies as well as the depth variations of the regular eclipses. In other words, the observed rates and directions of the dynamically forced apsidal motion (see Fig.~\ref{fig:etv}) and orbital precession, as well as the shape and amplitude of the medium-period dynamical perturbations in the motion of the innermost EB (see Fig.~\ref{fig:etv}), are in accord only with the flat, prograde configuration, and hence, a retrograde scenario can be ruled out with high confidence.

Turning to the Sector 40 \textit{TESS} light curve (Fig.~\ref{fig:model40}) a careful inspection reveals that the difference between the primary and secondary eclipse depths in the model solution is significantly larger than what is observed. This suggests that the surface brightness and, hence, temperature ratio of stars Ab and Aa should be much closer to unity than what was obtained in our \texttt{PARSEC} isochrone-constrained astrophysical model ($T_\mathrm{Ab}/T_\mathrm{Aa}=0.960\pm0.002$). This temperature ratio in our model basically depends on the mass ratio of the inner pair ($q_1=0.784\pm0.002$). The value of this latter quantity, however, is extremely robust in our model, especially due to the precise shapes and timings of the extra outer eclipses. These are very sensitive to the sizes of the physical orbits of stars Aa and Ab relative to each other, the latter of which is determined via their mass ratio. Hence, we conclude, that one or both of the very low mass components of the innermost binary do not strictly follow the theoretical mass-luminosity relation at the lowest mass end of the main sequence.

\begin{figure*}[h]
    \centering
    \includegraphics[width=1.0\linewidth]{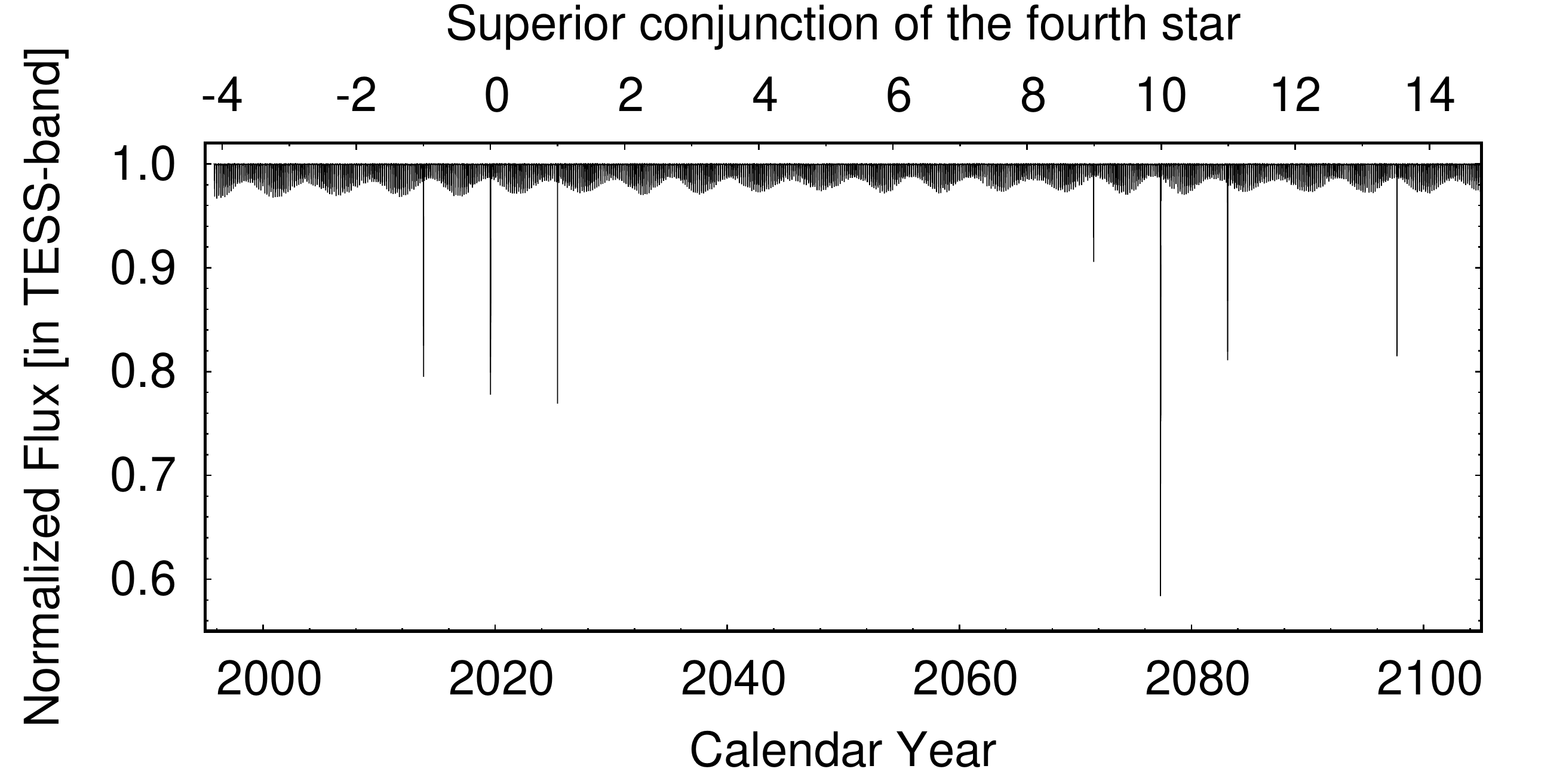}
    \caption{Model light curve of TIC 114936199 for the 21st century.}
   \label{fig:lc100yr}
\end{figure*}

In order to investigate the possibility for observing similarly extreme extra eclipses in the system, we modeled the orbital motion and hence, the light curve of TIC~114936199 for the entire 21st century. For input parameters we used our best-fit solution. Note, however, that as the outer orbital period is known only to within an uncertainty of 2-3 weeks, the results to be presented below serve as qualitative illustrations rather than precise predictions. With this caveat in mind, we plot the entire 21st century light curve of the system in Figure \ref{fig:lc100yr}. A first glance at this plot reveals that extra eclipses do not occur for all the superior conjunctions of star C. In addition to the extra eclipses during the 2019 superior conjunction in {\em TESS} sector 14, there may have been extra eclipses during the previous (2013) superior conjunction, and we might expect them during the forthcoming (2025) superior conjunction. Then, according to the present model, no further extra eclipses will occur in the epochs of the next 7 superior conjunctions until 2071, when, again, three superior conjunctions in a row will possibly exhibit extra eclipses. The last extra eclipse of the 21st century is expected in 2097, during an inferior conjunction of star C.

\begin{figure*}[h]
    \centering
    \includegraphics[width=1.0\linewidth]{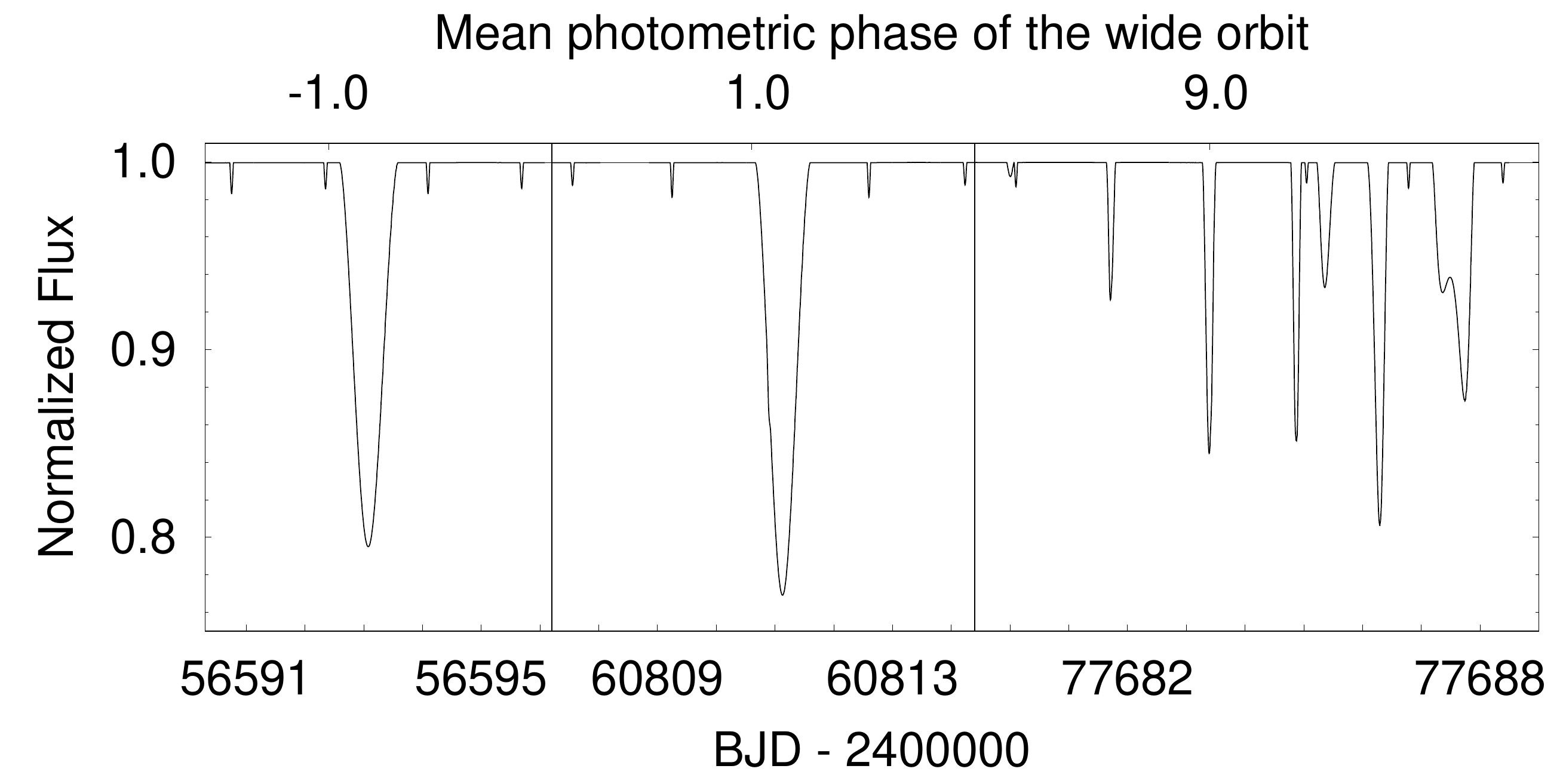}
    \includegraphics[width=1.0\linewidth]{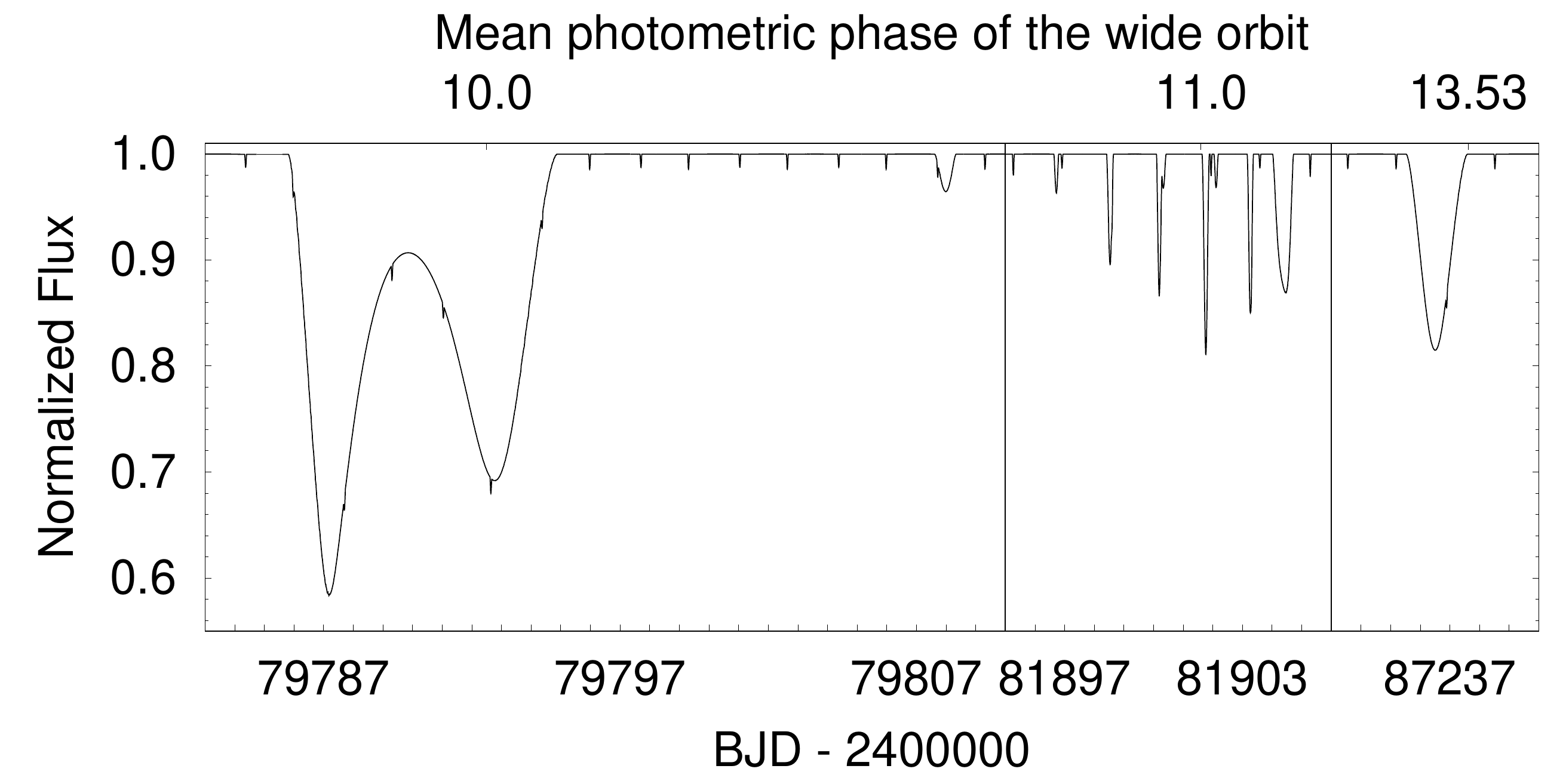}
    \caption{Additional extra eclipses in the 21st century. Vertical lines indicate time breaks.}
   \label{fig:lc100yrE3}
\end{figure*}

We zoom into all the additional six extra eclipses in Figure ~\ref{fig:lc100yrE3}. As one can see, the eclipses to occur during superior conjunctions 9 and 11 resemble the \textit{TESS}-observed 2019 event. In stark contrast, the outer eclipsing event prior to, and following, the {\em TESS}-observed event shows only one extra eclipse. The most spectacular event, however, is the one projected to occur during superior conjunction 10 (i.e., in 2077). In this case, instead of the inner pair, star B will eclipse star C. The primary event around the exact superior conjunction of star C will take $\sim$8 days, and will produce the deepest fading at least in the 21st century. Then, $\sim$three-weeks later, after a half revolution of star B on its orbit around the center of mass of the innermost binary, an additional, shallow dip will also be observable. Finally, in the case of the last extra eclipse of the present century, as it was mentioned above, the role of the latter two stars will be exchanged, and during this event star C will eclipse star B. Of course, we stress again, that due to the large uncertainties in the parameters of the outer orbit, these findings should be considered only as qualitative illustrations. Long-term targeted spectroscopic observations of the system would be very helpful to narrow the uncertainties of the parameters and, therefore, to obtain more reliable predictions for the forthcoming conjunctions.

\section{Spectral Energy Distribution}
\label{sec:sed}

\begin{figure}
    \centering
    \includegraphics[width=0.99\columnwidth]{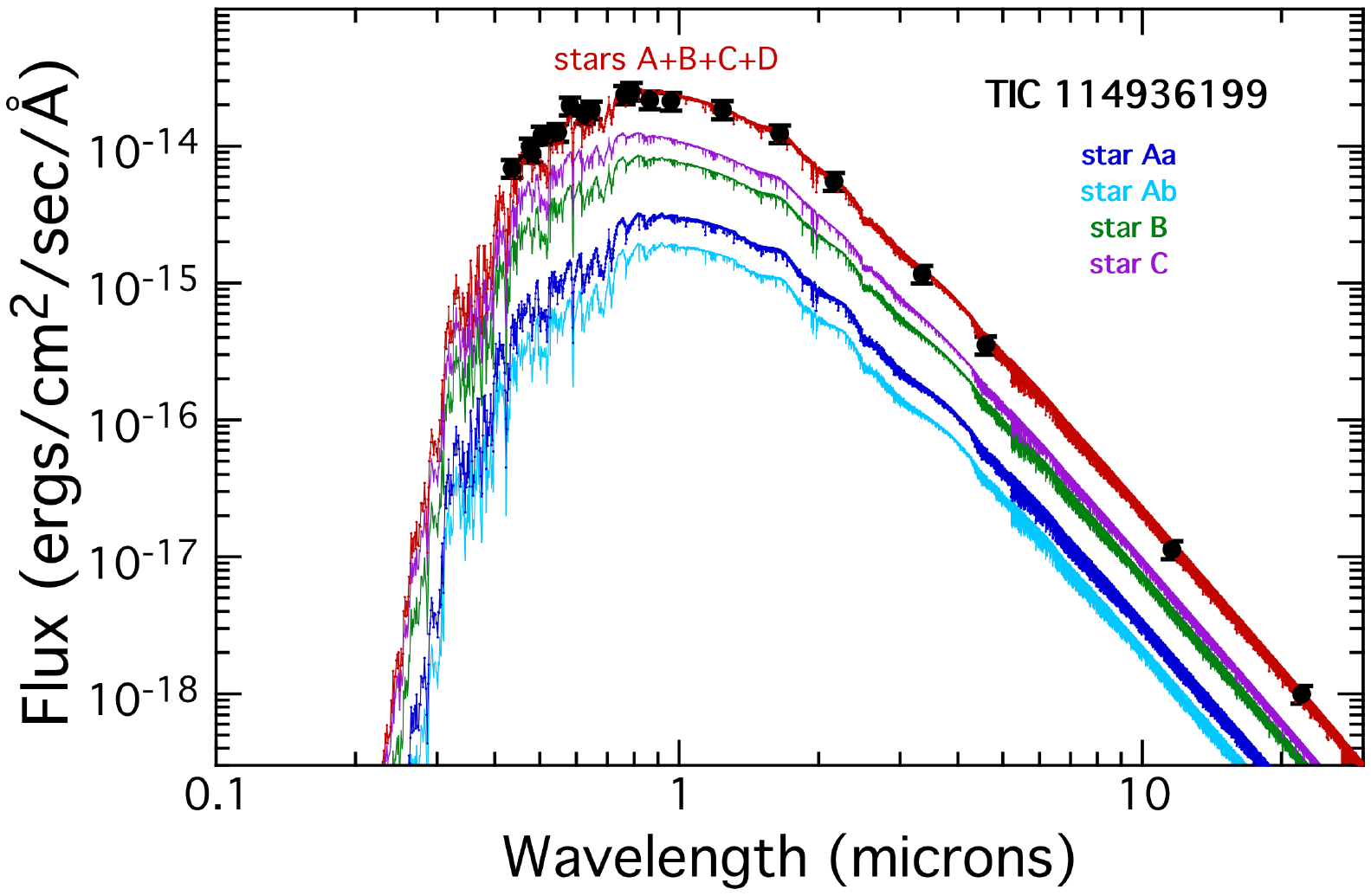}
    \caption{SED fit to 19 spectral fluxes that have been archived for TIC 114936199 
    \citep{2000A&AS..143...23O}.  The model curves, color coded for the contributions from the individual stars and the total flux, are from BT-Settl stellar atmospheres models \citep{2014IAUS..299..271A} of solar composition.}
   \label{fig:sed}
\end{figure}

Within the photodynamical fit, we utilized PARSEC-generated passband magnitudes which were fit to the observed magnitudes in different filters.  Here, to present the fit in a different format and as a sanity check, we show in Fig.~\ref{fig:sed} the 19 observed spectral fluxes from VizieR \citep{2000A&AS..143...23O} overplotted with model BT-Settl spectra \citep{2014IAUS..299..271A}.  The blue, cyan, green and purple curves represent the individual contributions of stars Ab, Ab, B, and C, respectively. The red curve is the sum of the individual stellar contributions to the total flux.  The fluxes of the BT-Settl model curves were integrated over the various filter bands used to make the measurements, and those were used to compute $\chi^2$.

\section{Discussion}
\label{sec:discussion}

The TIC 114936199 system is some 500 Myr old (see Section \ref{sec:mcmc}).  One question that arises in this context is whether the system can be expected to be dynamically stable going forward.  In this regard, we note that the inner triple is ``tight'' in the sense that $P_{\rm triple}/P_{\rm eb} \simeq 15$, but that still exceeds the minimum stable period ratio for the given masses and outer eccentricity by a factor of $\sim$3 based on the expressions of \citet{2001MNRAS.321..398M} and \citet{2008msah.conf...11M}.   Lengthy numerical integrations of such systems are impractical for billions of years, but we are confident that the inner triple is stable over long intervals (comparable to a Hubble time).  And, given that the outer orbit has $P_{\rm quad}/P_{\rm trip} \simeq 36$, this is even more stable, and thus the whole system should be long-term stable.

The most massive star in the system, star C, has a mass of 0.615 M$_\odot$.  For a solar composition such stars require 40 Gyr to evolve to the point of ascending the giant branch.  Since that is much longer than the current age of the Universe, there is not much point in speculating about what will happen when star C ultimately expands and overfills its Roche lobe.  The three lower mass stars in the system will hardly budge off the zero-age main sequence in the foreseeable future.

In terms of possible formation scenarios for this system we note that star B has a mass such that $M_{\rm Aa} < M_{\rm B} < M_{\rm Aa}+M_{\rm Ab}$, while star C has a mass in the range of $M_{\rm A} \approx M_{\rm C} < M_{\rm A}+M_{\rm B}$. These mass ratios fit in nicely with an accretion-based formation scenario \citep{2021Univ....7..352T}.

TIC 114936199 has the second tightest outer orbit among the few 2+1+1 type quadruple systems with known orbits. Moreover, it resembles the tightest such system, HIP 41431 \citep{2019MNRAS.487.4631B}, in several ways. HIP 41431 is also relatively flat, and formed by four red dwarf stars orbiting with periods of $\sim$2.9, $\sim$59, and $\sim$1441 days. The nearly coplanar orbits, and the small eccentricities liken both quadruples to multi-planet systems. On the other hand, the two systems are dissimilar in mass ratio as, in contrast to TIC 114936199, the outermost component of HIP 41431 has the lowest mass, while the inner triple subsystem consists of three similarly massive stars (about 0.61-0.63 $M_{\sun}$). Another apparent difference comes also from the less fortuitous orientation of HIP 41431. While the innermost pair of HIP 41431 is also an eclipsing binary, no outer orbit eclipse could be observed in that system.  Therefore, the mutual inclinations of the two wider orbits are known only with much higher uncertainties.

In this regard, we note that we were extremely lucky to have witnessed the very long-duration set of outer eclipses with {\it TESS}, both in terms of viewing the system at the proper outer orbital inclination angle and within the correct {\it TESS} sector of only a month.  Furthermore, we observed the system at a highly fortuitous time when the AB triple had just the correct orbital phase that is needed to slow the motion of star C on the sky relative to the A binary (i.e., near the conjunction of both the triple and the quadruple orbits).  As shown by the analysis in Section \ref{sec:mcmc}, the majority of superior conjunctions will not exhibit eclipses and, of the minority that do, the presence of such a rich and astrophysically useful eclipse pattern is unlikely to be exhibited.  As we demonstrated throughout this paper, deducing an orbital solution from the {\em TESS} sector 14 eclipse event was extremely difficult.  A much less informative eclipse, such as those shown in three of the six events of Figure \ref{fig:lc100yrE3}, would potentially not have been sufficient to allow for successful development of the model.

\section{A Likely Fifth Star in the System}
\label{sec:quintuple}

As this work on TIC 114936199 was nearing completion, we have become increasingly convinced by the steady improvement of the Gaia results that this quadruple also has a likely bound 5th star in the system.  Gaia source 4533878463909604224 is situated at an angular distance of 0.92$''$ from TIC 114936199 with a nearly identical parallax (well within the $1-\sigma$ uncertainties which are at the 1\% level), and 3 magnitudes fainter. Here we make some simple estimates of the mass and outermost orbit of this star which we tentatively call
``star D'' in its likely 2+1+1+1 configuration.  


The relative proper motions between TIC 114936199 and its fainter neighbor are $+11.4 \pm 0.12$ mas yr$^{-1}$ in RA and $+7.2 \pm 0.2$ mas yr$^{-1}$ in Dec. These correspond to physical velocities on the plane of the sky at the source of 6.96 km s$^{-1}$ in RA and $-4.59$ km s$^{-1}$ in Dec {\ron(8.33 km s$^{-1}$ in magnitude)}, with small formal uncertainties. At the current minimum separation between TIC 114936199 and star D of 120 AU, the escape speed from the quadruple's total mass of 1.8 M$_\odot$, would be 5.2 km s$^{-1}$.  However, we note that some of the observed proper motion is actually due to bound orbital motion within the AB-C binary.  We have carefully computed the mean speed on the sky of the center of light within the AB-C binary during the 34 months of the Gaia observations that went into DR3.  This comes out to be only 2.4 km s$^{-1}$.  Thus, if the orientation of the AB-C orbit on the sky were parallel to the PM vector (which is not certain) then this could lower the proper motion due to the ABC-D orbit to $\sim$5.9 km s$^{-1}$.  If so, the relative motions of the TIC 114936199 center of mass and star D would be such that star D would still not be quite bound to the quadruple, but perhaps close enough within the uncertainties involved.

On the other hand we would argue that (i) the nearly identical (to within 1\%) distances of TIC 114936199 and star D, (ii) their proximity in the sky (0.92$''$), as well as (iii) the similar low mass character of star D (see below) to the stars in TIC 114936199, all make it virtually certain that D is at least physically associated with TIC 114936199, if not actually gravitationally bound. One possibility is that Gaia's proper motion measurements have been distorted by the proximity of the two stars, and that they are, in fact, somewhat smaller than what is presented in Gaia DR3. The alternative is that the two objects (TIC 114936199 and star D) are not actually bound, but are simply approximately co-moving objects. However, at a total differential speed of $\simeq 8$ km s$^{-1}$, and an estimated age of $\sim$500 Myr (see Table 4) the two objects would have separated by $\sim$4000 pc at the current epoch. Since the distance to both objects is only 130 pc, and they are currently separated on the sky plane by 120 AU, this scenario seems implausible.


Therefore we tentatively conclude that (i) TIC 114936199 and star D are indeed gravitationally bound, (ii) the Gaia relative proper motions are off by about a factor of 2, and (iii) this is actually a quintuple system with a 2+1+1+1 configuration.  

Given that star D has about 5.4\% of the light of TIC 114936199 in the Gaia band, and that stars Aa and Ab contribute 8.7\% and 4.6\% of the system light, respectively, the mass of star D lies between the masses of these stars of 0.382 M$_\odot$ and 0.300 M$_\odot$ (see Table 4).  A simple interpolation would suggest a mass for star D of $\simeq 0.32$ M$_\odot$.

Finally, if we take the size of the D orbit about TIC 114936199 to be approximately 120 AU, and we adopt the mass of TIC 114936199 from Table 4, and use the mass estimate for D above, then the outermost period in this systems could well be around 900 years. The sequential period ratios of the four hierarchical orbits would then be 1:15:41:150. Thus, this is overall a fairly tight quintuple.

\section{Summary}
\label{sec:summary}

We have presented the discovery of TIC 114936199 as a remarkable 2+1+1 quadruple star system eclipsing on the outer orbit, and we provided a unique physical model of the system, the first of its kind.  

We demonstrated how, using a single complex eclipsing event, we applied clues provided by the eclipses to hypothesize the nature of the eclipse and the configuration of the system.  Then, by broadly estimating orbital periods and stellar masses, and using a combination of heuristic optimization algorithms and MCMC modeling, we were able to find a solution to the system.

Our solution is very robust, beautifully fitting the {\em TESS} light curve over three different sectors of observation spanning several years, independent ground-based observations, the ETV curves, the TRES RV measurements, and the net SED of the system.

We have provided qualitative predictions of future eclipse events with the caveat that the outer orbital period is known to $\sim$1\%, which is 
only marginally sufficient to project ahead by multiple outer orbital cycles.  Dedicated spectral observation of TIC 114936199 may provide the necessary radial velocity measurements to further constrain the solution and determine a narrow range for the outer orbit, allowing for observation of future outer eclipse events.

We tentatively conclude that TIC 11436199 is actually part of a larger quintuple system in a 2+1+1+1 configuration.  This outer star likely has a mass of $\sim$0.32 M$_\odot$ and an orbital period of $\sim$900 years.

We believe that the methods outlined in this paper could be applied to KIC 5255552 and KIC 285696, to determine if a 2+2 or 2+1+1 quadruple model could provide an adequate solution to either system, which we suspect it will.

\section*{acknowledgments}

Resources supporting this work were provided by the NASA High-End Computing (HEC) Program through the NASA Center for Climate Simulation (NCCS) at Goddard Space Flight Center.  

This paper includes data collected by the {\em TESS} mission, which are publicly available from the Mikulski Archive for Space Telescopes (MAST). Funding for the {\em TESS} mission is provided by NASA's Science Mission directorate.

We thank P.\ Berlind, M.\ Calkins, and G.\ Esquerdo for their help in obtaining the spectroscopic observations of TIC 114936199 with TRES.

We thank Petr Zasche for reduction of the data from the observation obtained by HK.

ZG acknowledges the support of the Hungarian National Research, Development and Innovation Office (NKFIH) grant K-125015, the PRODEX Experiment Agreement No. 4000137122 between the ELTE Eötvös Loránd University and the European Space Agency (ESA-D/SCI-LE-2021-0025), the VEGA grant of the Slovak Academy of Sciences No. 2/0031/22, the Slovak Research and Development Agency contract No. APVV-20-0148, and the support of the city of Szombathely.

TP was supported by the VEGA grant of the Slovak Academy of Sciences number 2/0031/22 and the Slovak Research and Development Agency under contract No. APVV-20-0148.

This research has made use of the Exoplanet Follow-up Observation Program website, which is operated by the California Institute of Technology, under contract with the National Aeronautics and Space Administration under the Exoplanet Exploration Program. 

\facilities{
\emph{Gaia},
NCCS,
MAST,
{\em TESS}}

\software{
{\tt Astrocut} \citep{astrocut},
{\tt Astropy} \citep{astropy2013,astropy2018}, 
{\tt Eleanor} \citep{eleanor},
{\tt IPython} \citep{ipython},
{\tt Keras} \citep{keras},
{\tt LcTools} \citep{2021arXiv210310285S},
{\tt Lightkurve} \citep{lightkurve},
{\tt Matplotlib} \citep{matplotlib},
{\tt Mpi4py} \citep{mpi4py2008},
{\tt NumPy} \citep{numpy}, 
{\tt Pandas} \citep{pandas},
{\tt PySwarms} \citep{pyswarmsJOSS2018},
{\tt Rebound} \citep{2012A&A...537A.128R},
{\tt Scikit-learn} \citep{scikit-learn},
{\tt SciPy} \citep{scipy},
{\tt Shapely} \citep{shapely},
{\tt Tensorflow} \citep{tensorflow},
{\tt Tess-point} \citep{tess-point}
}

\bibliography{refs}{}
\bibliographystyle{aasjournal}

\end{document}